\documentclass[apj]{emulateapj}
\usepackage{natbib,graphicx,subfigure,url}
\usepackage[usenames]{color}
\usepackage{multirow}
\shorttitle{Quiescent Black Hole X-ray Binaries}
\shortauthors{Plotkin, Gallo \& Jonker}

\begin{document}

\title{The X-ray Spectral Evolution of Galactic Black Hole X-ray Binaries Toward Quiescence}
\shorttitle{Quiescent Black Hole X-ray Binaries}

\author{
Richard.~M.~Plotkin,\altaffilmark{1} 
Elena~Gallo,\altaffilmark{1}
and Peter~G.~Jonker\altaffilmark{2,3,4}
}

\altaffiltext{1}{Department of Astronomy, University of Michigan, 500 Church St., Ann Arbor, MI 48109, USA; rplotkin@umich.edu}
\altaffiltext{2}{SRON, Netherlands Institute for Space Research, Sorbonnelaan 2, 3584 CA, Utrecht, the Netherlands}
\altaffiltext{3}{Harvard-Smithsonian Center for Astrophysics, 60 Garden Street, Cambridge, MA 02138, USA}
\altaffiltext{4}{Department of Astrophysics/IMAPP, Radboud University Nijmegen, PO Box 9010, 6500 GL, Nijmegen, the Netherlands}

\newcommand{\hxvii}{H~1743-322}   
\newcommand{\maxi}{MAXI~J1659-152}
\newcommand{\jxvii}{XTE~J1752-223}   
\newcommand{\jxix}{XTE~J1908+094}  
\newcommand{\avi}{A0620-00}
\newcommand{\gro}{GRO~J1655-40}
\newcommand{\gx}{GX~339-4}
\newcommand{\viv}{V404~Cyg}
\newcommand{\jxi}{XTE~J1118+480}
\newcommand{\jxv}{XTE~J1550-564}
\newcommand{\jxvi}{XTE~J1650-500}

\newcommand{\nh}{$N_{\rm H}$}    
\newcommand{\lledd}{$L_X/L_{\rm Edd}$}  
\newcommand{\ledd}{$L_{\rm Edd}$}   

\newcommand{\xrb}{BHXB}
\newcommand{\mdot}{$\dot{m}$}
\newcommand{\Mdot}{$\dot{M}$}

\newcommand{\nar}{New A Rev.}

\begin{abstract}
Most transient black hole X-ray binaries (\xrb s) spend the bulk of their time in a quiescent state, where they  accrete matter  from their companion star at highly sub-Eddington luminosities (we define quiescence here as a normalized Eddington ratio $l_x = L_{0.5-10~{\rm keV}}/L_{Edd} < 10^{-5}$).   Here, we present \textit{Chandra} X-ray imaging spectroscopy for three \xrb\ systems (\hxvii, \maxi, and \jxvii) as they fade into quiescence following an outburst.     Multiple X-ray observations were taken within one month of each other, allowing us to track each individual system's X-ray spectral evolution during its decay.   We  compare these three systems to  other  \xrb\ systems.  We confirm that  quiescent \xrb s have softer  X-ray spectra than low-hard state \xrb s, and that quiescent \xrb\ spectral properties show no dependence on the binary system's orbital parameters.  However,  the observed anti-correlation between X-ray photon index ($\Gamma$) and $l_x$  in the low-hard state does not continue once a \xrb\ enters quiescence.    Instead, $\Gamma$ plateaus to an average $\langle \Gamma \rangle = 2.08 \pm 0.07$ by the time $l_x$ reaches $\sim$10$^{-5}$.   $l_x \sim 10^{-5}$ is thus an observationally-motivated upper limit for the beginning of the quiescent spectral state.    Our results are discussed in the context of different accretion flow models, and across the black hole mass scale.

\end{abstract}


\section{Introduction} 
\label{sec:intro}

 Black holes are common in the Universe, with a supermassive black hole likely at the center of every large galaxy \citep[e.g.,][]{kormendy95, marconi04},  and predictions of upwards of 10$^8 - 10^9$  stellar mass black holes in the Milky Way alone \citep[e.g.,][]{shapiro83, maccarone05, fender13}.   Yet, we can ultimately observe just a small fraction of these black holes, and it is only for an even smaller subset that we can appeal to dynamical interactions to infer their properties.  We are forced to study the vast majority of known black holes  through indirect methods,  like the radiative signatures produced  when they accrete matter.   There is thus strong motivation to better understand  accretion onto black holes.   Studying accreting black holes not only provides direct constraints on Galactic black hole X-ray binaries (\xrb s), Active Galactic Nuclei (AGNs), and black hole feedback, but also broad insight into many other classes of objects where similar physics is at play (e.g., young stars, white dwarfs, neutron stars, gamma-ray bursts; see e.g., \citealt{meier01a, migliari06, kording08, scaringi12}).   
 
  X-ray emission is a universal, but not yet fully understood, feature of  accreting black holes.  Complicating matters is that the most important physical mechanism(s) responsible for X-ray emission  seems to depend largely (albeit not entirely) on the normalized mass accretion rate \mdot~$\equiv \dot{M}/\dot{M}_{Edd}$ \citep[e.g.,][]{esin97, trump11}, where $\dot{M}$ is the mass accretion rate in physical units (g~s$^{-1}$) and $\dot{M}_{Edd}$ is the Eddington mass accretion limit.  Throughout this paper, we  use the normalized Eddington X-ray luminosity ($l_x = L_X/L_{\rm Edd}$; where $L_X$ is the X-ray luminosity from 0.5--10~keV and \ledd~$= 1.26 \times 10^{38} \left[M/M_\odot\right]$~erg~s$^{-1}$ for ionized Hydrogen) as a proxy for \mdot.  We note that  $l_x$ and \mdot\ correlate with each other, but in general $ l_x \neq$~\mdot.      Very low-\mdot\ represents a particularly important regime.  Most transient \xrb s spend the bulk of their time in a very weakly accreting quiescent state.  While there is not a standard definition for quiescence, a commonly used criterion is $L_X \sim 10^{30.5} - 10^{33.5}$~erg~s$^{-1}$ (corresponding to $l_x \sim 10^{-8.5} - 10^{-5.5}$ for a 10~$M_\odot$ black hole; \citealt{remillard06}). Most supermassive black holes in the local Universe also accrete just as weakly.   Yet, we still need a better understanding on  how quiescent black holes produce high-energy radiation, and on the structure and geometry of their accretion flows.
  
 When a power law is fit to the X-ray spectra of black holes with $l_x \la 10^{-2}$,  an anti-correlation is seen between $l_x$ and the photon power-law index\footnote{The X-ray photon index $\Gamma$ is defined as $N(E) = N_0 \left(E/E_0\right)^{-\Gamma}$, where $N(E)$ is the number of photons at a given energy $E$, $N_0$ is the photon number normalization, and $E_0$ is the reference energy (we set $E_0$  to 1~keV here).} 
$\Gamma$  \citep[e.g.,][]{yuan07, wu08, constantin09, gu09, sobolewska11a, younes11, gultekin12}.   That is,  lower-accretion rate black holes have softer (i.e., steeper) X-ray spectra.\footnote{This $\Gamma - l_x$ anti-correlation is opposite to the trend observed at higher accretion rates for both \xrb s and for luminous quasars \citep[e.g.,][]{kubota04, shemmer08, grupe10}.}   
 However, this anti-correlation has not yet been fully probed below $l_x \sim10^{-4}$.   Exploring the $\Gamma - l_x$ anti-correlation down to lower $l_x$ will   lead to a  better understanding of the properties of black hole accretion flows in quiescence.    Whether the X-ray spectral softening continues  at all $l_x$ or eventually saturates, and also if the softening is gradual or abrupt, can provide constraints on accretion disk/jet models \citep[e.g.,][]{tomsick01}.    Furthermore, characterizing the behavior of $\Gamma$ at very-low $l_x$ may  allow for a more systematically-determined definition of the quiescent state.

In this paper, we consider  observations of \xrb s in order to help constrain this relatively unexplored very low-$l_x$ parameter space ($l_X \la 10^{-4}$).   Transient \xrb s undergo outbursts in luminosity that are marked by X-ray spectral state transitions accompanied by characteristic variability properties, as well as outflows in the forms of jets and/or winds \citep[][also see, e.g., \citealt{fender04, fender05, fender09, belloni05, homan05} for reviews on outburst phenomenology]{remillard06}.  Here, we consider only the final parts of \xrb\ outbursts, after a \xrb\ transitions back into a hard X-ray state (with $\Gamma \sim 1.5$; i.e., the ``low-hard'' state) and then  decays  into quiescence.     So far, around a dozen \xrb s have been observed in quiescence with high-enough sensitivity to extract X-ray spectral information.   These quiescent \xrb s tend to be relatively soft with $\Gamma \sim 2$ (e.g., \citealt{ebisawa94, tomsick04, corbel06}).   However, the majority of the observed \xrb\ systems only have one or two X-ray observations in quiescence, usually separated by years.  Thus, it is difficult to provide definitive statements on the  details of how the X-ray spectral softening occurs on a case-by-base basis.
    
We initially focus on archival \textit{Chandra} observations of three  \xrb\ systems (\hxvii, \maxi, and \jxvii).    The observations were taken as part of joint \textit{Chandra} and Very Large Array (VLA)  target of opportunity (TOO) programs with PI Jonker.   These Jonker et al.\ TOO programs were approved over several \textit{Chandra} cycles to monitor individual \xrb\ systems at the tail of end an outburst (\hxvii\ was observed during \textit{Chandra} cycle-9, \jxvii\ during cycle-11, and \maxi\ during cycle-12).  The observations were triggered once a radio counterpart was identified and the X-ray flux dropped below a certain threshold  (usually $10^{-11}-10^{-10}$~erg~s$^{-1}$~cm$^{-2}$ from 0.5--10~keV, which typically  corresponds to $l_x < 10^{-4}$), with a preference toward observing systems with low line of sight Hydrogen column densities.   Usually 4--8 (nearly) simultaneous \textit{Chandra} X-ray and VLA radio observations were taken within 30 days of triggering.     The \textit{Chandra} observations were designed to obtain enough counts to allow for imaging spectroscopy, making an archival study based on these TOO programs ideal for studying the X-ray spectral evolution of transient \xrb s as they fade into quiescence, as both a function of time and of luminosity.  

The data for \hxvii, \maxi, and \jxvii\ were originally published in \citet{jonker10}, \citet{jonker12}, and \citet{ratti12}, respectively.   These publications  focus on the evolution of each source's radio/X-ray luminosity during the fade into quiescence, while here we instead focus on their X-ray spectral evolution.   After tracking the spectral evolution for the above individual systems, we then  compare their spectral properties  to other \xrb\ systems with available  X-ray spectra {at $l_x < 10^{-4}$ in the literature.  Simultaneously examining all available X-ray spectra  provides new insight into how (and at what X-ray luminosity) black holes transition into quiescence, and whether or not quiescence is simply an extension of the low-hard state.  Our sample and data reduction are described in \S \ref{sec:obs}; the best-fit X-ray spectral properties are presented in \S \ref{sec:results}, and our results are discussed and summarized in \S \ref{sec:disc} and \S \ref{sec:conc}, respectively.      All reported X-ray luminosities are unabsorbed (i.e., corrected for the effects of extinction) and calculated from 0.5 -- 10~keV by integrating  over each observation's best-fit  power law spectrum, and all quoted measurement uncertainties are at the 68\% level (i.e., $\Delta \chi^2 = 1.0$ for one parameter of interest).

\section{Chandra X-ray Observations}
\label{sec:obs}
\subsection{\textit{Chandra} Data Reductions and Spectral Fitting}
\label{sec:reduction}
There are a total of 18 unique \textit{Chandra} observations covering \hxvii\ (eight observations taken in 2008 March), \maxi\ (six observations taken from 2011 April -- October), and \jxvii\ (four observations taken from 2010 July -- August).  Each source's properties are given in Table~\ref{tab:srcinfo}, and the \textit{Chandra} observations (including the obsIDs)  are summarized in Table~\ref{tab:obslog}.  We note that Jonker et al.\  carried out  similar joint \textit{Chandra}/VLA TOO programs for two other sources, \jxix\  (cycle-4) and V4641~Sgr (cycle-5).  However, only three X-ray observations were taken for \jxix.  Only one of these three observations has $l_x < 10^{-4}$, and that observation has  too few counts to fit an  X-ray spectrum \citep[see][]{jonker04}.  We thus do not consider \jxix\ here.    V4641~Sgr typically shows an unusually hard X-ray spectrum while in outburst  \citep[e.g.,][]{maitra06}, and this hard spectrum apparently persists even in our quiescent \textit{Chandra} observations.  Given its odd behavior, we do not consider V4641~Sgr here and we defer a discussion on V4641~Sgr in quiescence to a future paper (Gallo, Plotkin \& Jonker in prep).  

We re-analyze all 18 observations, to ensure that our data reductions are as uniform as possible from source to source, and that the latest calibration is applied to each observation.     The target is always placed on the back-illuminated S3  chip of the Advanced CCD Imaging Spectrometer (ACIS) detector \citep{garmire03}.   To help avoid photon pileup, the ACIS-S3 CCD is windowed to read out only the 1/8 chip subarray, except for four observations of \hxvii\ (obsIDs 8990, 9837, 9838, and 9839) that use the 1/2 chip subarray.  We reprocessed and analyzed all data with the {\tt CIAO}4.4 software \citep{fruscione06} developed by the \textit{Chandra} X-ray Center, employing the latest calibration files from  data base version 4.5.3.    We used the {\tt chandra\_repro} script to reprocess the raw data, and we created custom bad pixel maps for each observation.  Six of our longest exposures (obsIDs 8990, 9837, 9838, 9839, 11056, and 12443) were taken in {\tt VFAINT}  data mode.  For these observations, pulse height information in a 5 x 5 pixel region (instead of a 3 x 3 pixel region) is telemetered down, allowing for a more rigorous cleaning of background events (e.g., caused by cosmic rays, etc.). Unless otherwise stated, we only consider events with photon energies between 0.5--7~keV to minimize the ACIS high-energy particle background.   All data are used, since we do not see any evidence for background flares.  All 18 \textit{Chandra} observations contain a detection of the target source, typically with  tens to hundreds of counts (see Table \ref{tab:obslog}). 

We extract source counts and spectra within circular apertures centered on accurately known sky positions for each source (see Table~\ref{tab:srcinfo}).  We generally use 2-arcsec radius extraction regions.  However, in order to enclose a sufficiently high number of total source counts, we adopt 3-arcsec radius apertures for two observations of \hxvii\ (obsIDs  8987 and 8988) and for one observation of \jxvii\ (obsID 12310).  We similarly adopt a 5-arcsec radius aperture for one observation of \maxi\ (obsID 12441).  Finally, our longest exposures for \jxvii\ (obsIDs 11055 and 11056) reveal faint sources very close to the source position.\footnote{Some of these faint sources appear to be extended X-ray jet emission \citep{ratti12}.} 
  We thus use 1.5-arcsec radius apertures for these two observations. To extract sky counts, we use a circular annulus  with a 10-arcsec inner radius and a 20-arcsec outer radius whenever possible.   However, some observations have faint  sources within these radii.  So, for all four observations of \jxvii\ (obsIDs 11053, 11055, 11056, and 12310) we use a 12-arcsec inner radius and an 18-arcsec outer radius; for our two longest exposures of  \maxi\ (obsIDs 12442 and 12443) we use a 6-arcsec inner radius and a 15-arcsec outer radius.   Since one observation of \maxi\ (obsID 12441) shows readout streaks related to  its high count rate (see \S \ref{sec:pileup}),   we extract sky counts for this  observation using a 15-arcsec radius circular aperture centered on a source-free region of the CCD near the target source.  This off-source aperture avoids source photons from the streak potentially contaminating our sky background.

X-ray spectra are extracted using {\tt specextract} in {\tt CIAO}.    We create background and source response matrix files and  auxiliary response files for each observation, applying  an energy dependent point-source aperture correction to the latter to account for the fraction of enclosed energy within our adopted source apertures.   Spectra are then fit using {\tt ISIS} version 1.6.2-10 \citep{houck00}.  We include background counts in our fits using Cash statistics \citep{cash79},\footnote{The Cash statistic converges toward $\chi^{2}$ when the number of source counts is large.}    
 adopting an absorbed power law model for all sources (i.e., {\tt phabs*powerlaw}).\footnote{For the photoelectric absorption model, we use cross sections from \citet{balucinska-church92} with updated He cross-sections from \citet{yan98}, and we use abundances from \citet{anders89}. } 
   We only allow the power law normalization and photon index ($\Gamma$) to vary as free parameters, fixing the Hydrogen column density \nh\ to the same values adopted in our previous work on these sources, as listed in Table~\ref{tab:srcinfo}.  An absorbed powerlaw model provides adequate fits, but we note that there are likely other models that could fit equally well.   Our primary goal  is to compare the spectral shape of several  sources at  different $l_x$.  Thus, especially given the relatively low number of source counts in most of our observations, an absorbed powerlaw serves as the simplest way to  parameterize the X-ray spectral shape, and to then uniformly compare all of our data.   Our best fit spectral parameters are summarized in Table~\ref{tab:fits}, and we note that our best-fit parameters are consistent with the previously published values based on these data.  
  
 There are three observations with too few source counts to obtain adequate spectral fits, including one observation for \hxvii\ (obsID 9833), and two observations for \maxi\ (obsIDs 12440 and 12442).   We instead estimate effective photon indices for these three observations using their observed source count rates in  soft (0.5-2.0~keV) and hard (2.0-7.0~keV) X-ray bands.  Aperture corrections are applied to each band's count rates to account for the  enclosed energy within each circular extraction region ($\sim$2~arcsec radius apertures enclose an average of $\sim$95\ and 88\% within the soft and hard bands, respectively).   We then use the  \textit{Chandra }Portable, Interactive Multi-Mission Simulator \citep[PIMMS;][]{mukai93}\footnote{\url{http://cxc.harvard.edu/toolkit/pimms.jsp}} 
   to infer effective photon indices, applying the appropriate effective area curves  for the \textit{Chandra} cycle when each observation was taken.     We also calculate effective photon indices for the 14  observations that have good spectral fits and that do not display signs of photon pileup.  By comparing these 14 effective photon indices to their corresponding  best-fit (spectroscopic) photon indices,  we  estimate that our effective photon indices are accurate to approximately $\pm$0.5 (the maximum  difference between the two types of photon index measures).

\subsubsection{Photon Pileup}
\label{sec:pileup}
One observation of \maxi\ (obsID 12441) has a sufficiently high enough count rate to suffer from the effects of photon pileup.  Here, two or more photons could arrive within a CCD detector region during a single frame time integration (0.4 s for this observation) and  subsequently be registered as a single event.   We  fit this  dataset in {\tt ISIS} using the pileup model of \citet{davis01}.    When fitting with the pileup model, we consider all events with energies larger than 0.5~keV.\footnote{Filtering the data from 0.5--7~keV as in the previous section may provide less reliable fits with the pileup model (see \url{http://cxc.harvard.edu/ciao/why/filter\_energy.html}).} 
We find the pileup is rather mild, with a pileup fraction $f_{\rm pile}=0.021$.  This observation also shows a readout streak, which is caused by photons that are collected during the 41 ms it takes to transfer the image into the readout buffer.  This streak is not affected by pileup, and it  contains sufficient counts that we can  extract a source spectrum from the streak photons.    Thus, we also extract a streak spectrum, using  two  $\sim$3 $\times$ 25 arcsec$^{2}$ boxes located $\sim$5~arcsec northeast and southwest of the source position, and we extract sky counts from surrounding regions of size $\sim$60 $\times$ 25 arcsec$^{2}$ centered on each source extraction box (and  excluding streak photons).    There are a total of 201 streak source photons.

The effective exposure time for the readout streak data is much lower than the total exposure time, since for each frame time integration of 0.4 sec each streak pixel is only exposed for 40 $\mu$s (the time to transfer and read  one pixel).  Our streak extraction region contains approximately 100 streak pixels, yielding an effective exposure time of 0.004~s per frame.  With a total time on source of 18.1~ks, we have a combined 18100/0.4 = 45250 frames.  The effective streak exposure time is thus approximately 0.004 $\times$ 45250 = 181~s.    Finally, we create a response matrix file and an auxiliary response file at the location of \maxi, using {\tt mkacisrmf} and {\tt mkarf}, respectively (and we adjust the auxiliary response file for the 181~s effective exposure time).  Our best spectral fit  is consistent with the on-source extraction fit with the pileup model within 1$\sigma$, and we  include both fits in Table~\ref{tab:fits}.   Throughout this paper, we adopt the spectral parameters extracted from the readout streak in order to minimize the number of free parameters to our fit.  Our results are similar, however,  if we instead used the pileup model best-fit parameters.

\subsection{Chandra Spectroscopy of Other Weakly Accreting \xrb\ Systems}
\label{sec:litobs}
We wish to compare our observations that track three individual \xrb s as they fade into quiescence with as many other  systems as possible.    Unfortunately, no other system has sufficient X-ray temporal coverage  to track their spectral evolution on a case-by-case basis.  However, comparison to the ensemble of all other \xrb s observed at $l_x \la 10^{-4}$ will  allow additional insight into the average properties of quiescent \xrb s.    Only \textit{Chandra} and \textit{XMM-Newton} provide enough sensitivity to obtain X-ray spectra at these very low X-ray luminosities with reasonable exposure times.  Searching the \textit{Chandra} and \textit{XMM-Newton} archives, there are only around a dozen \xrb s with adequate observations.  The vast majority of observations are with \textit{Chandra} (using ACIS-S imaging spectroscopy), and every source with an \textit{XMM-Newton} spectrum at $l_x \la 10^{-4}$ also has \textit{Chandra} coverage at a similar X-ray luminosity.  Thus, in order to keep comparisons between systems as uniform as possible (and to minimize the potential for source confusion due to \textit{XMM-Newton}'s coarser spatial resolution), we decide to directly compare only to other \textit{Chandra} ACIS-S observations.   We  re-analyze all archival \textit{Chandra} observations and perform our own spectral fits in order to include the latest calibrations, and also to help alleviate potential systematics that could be introduced by simply combining a heterogeneous set of data reductions directly from the literature.

 We require archival \textit{Chandra}  observations to contain at least 20 source counts.     This constraint adds another 19  \textit{Chandra} observations divided over  7 additional  systems (two observations of \avi, two observations of \gro, two observations of \gx, two observations of \viv, one observation of \jxi,  seven observations of \jxv,\footnote{Although there are a total of seven observations for \jxv, these observations are separated by months to years with similar observed X-ray fluxes.  Thus, they do not provide the same opportunity to track X-ray spectral evolution as for \hxvii, \maxi, and \jxvii.} 
 and three observation of \jxvi).   We include basic properties for each source in Table~\ref{tab:srcinfo}, and a log of these 19 \textit{Chandra} X-ray observations can be found in Table~\ref{tab:litobslog}.  We  note that the observations were originally taken for primary science purposes different than ours, and we list the publication that first presented each \textit{Chandra} observation in Table~\ref{tab:litobslog}.    Combined with our  other three \xrb\ systems in Tables~\ref{tab:obslog} and \ref{tab:fits}, we have a total of 37 \textit{Chandra} observations covering 10 systems.  We note that we do not consider any system that has X-ray observations in quiescence but with too few counts to fit an X-ray spectrum \citep[e.g., many of the systems in Figure~4 of][]{reynolds11}.  Nor do we include any X-ray observations of systems at $l_x > 10^{-4}$, since that luminosity regime has already been well-explored in the literature \citep[e.g.,][]{wu08}.

\subsubsection{X-ray Data Reduction of Other \textit{Chandra} Observations}
\label{sec:litreduce}
We follow identical data reduction and spectral fitting procedures as described in \S \ref{sec:reduction}.   These archival \textit{Chandra} observations generally have a similar setup as described earlier.   However, most observations read out the entire CCD instead of a subarray except for the following: for \gx, obsID 12410 uses the 1/8 chip subarray; for \viv, obsID 97 uses the 1/4 chip subarray and obsID 3808 uses the 1/8 chip sub-array; and for \jxv, obsID 3448 uses the 1/4 chip-subarray.   \jxvi\ is fairly bright in all three observations.  As described in \citet{tomsick04}, to reduce pileup the first two observations (obsIDs 3400 and 3401) were taken with the High Energy Transmission Grating Spectrometer (HETGS) in place.  The third observation (obsID 2731) did not use the HETGS (since there were fewer counts per second), but to mitigate pileup the 1/8 chip subarray was used and \jxvi\ was also placed 2.7-arcmin off-axis to blur the point spread function.    For the HETGs observations, we reduce the zero-order images.  The effects of pileup are still present in all three observations of \jxvi, but none shows a readout streak with sufficient counts  to extract a spectrum.  

\begin{figure*}[htbp]
\center
\includegraphics[scale=0.6]{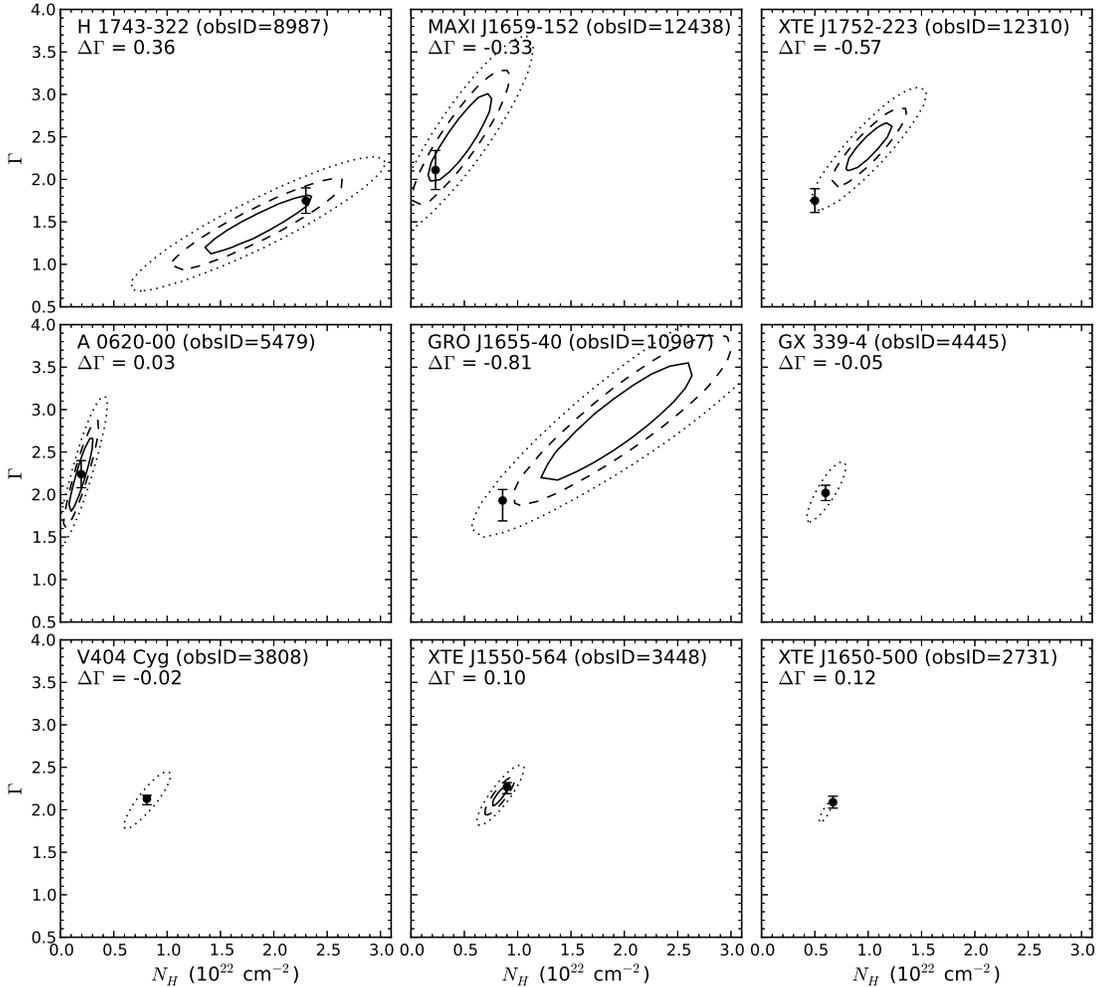}
\caption{Confidence contour maps of photon index ($\Gamma$) vs. Hydrogen column density (\nh), allowing \nh\ to vary as a free parameter for each system's \textit{Chandra} observation with the largest number of  counts (the second highest for \maxi\ to avoid pileup).  We omit \jxi\ because of its very small column density.  Solid, dashed, and dotted lines show changes in the Cash statistic ($\Delta C$) of 2.30, 4.61, and 9.21, respectively (i.e., the 68, 90, and 99\% confidence intervals for two parameters of interest). For clarity, we only draw the 99\%\ contour for \gx, \viv, and \jxvi.  The filled circles show the best-fit $\Gamma$ values when fixing \nh\ to the values adopted in the text (see Table~\ref{tab:srcinfo}).  The difference between the best-fit $\Gamma$ when fixing \nh\ and when allowing \nh\ to vary is reported as $\Delta \Gamma = \Gamma (N_{\rm H, fixed}) - \Gamma (N_{\rm H, free})$ in the top-left corner of each panel, in order to (roughly) quantify the systematic uncertainty that could be introduced  by fixing \nh\ to the values in  Table~\ref{tab:srcinfo}.}
\label{fig:conf}
\end{figure*}

 Most  of the 19 observations are taken in {\tt FAINT} data mode, except  obsIDs 3422 (\jxi) and 10097 (\gro) are taken in {\tt VFAINT} data mode.   We again extract source counts using 2-arcsec radius circular apertures (centered on the known source positions listed in Table~\ref{tab:srcinfo}), and we extract sky counts using circular annuli with inner and outer radii of 10- and 20-arcsec, respectively.  However, for \avi\ (obsIDs 95 and 5479), \gx\ (obsIDs 4445 and 12410), and \viv\  (obsIDs 97 and 3808) we instead extract source counts with a 3-arcsec radius circular aperture, since there are a relatively larger number of source counts in these observations.  For \avi, we extract sky counts using a circular annulus with 15-arcsec inner radius and 25-arcsec outer radius, in order to avoid nearby  sources.  \jxv\ has a relatively large number of source counts in obsID 3448 ($\sim$1200 photons); for this observation we thus extract source counts using a 5-arcsec radius circular aperture.     \jxv\ is also in a particularly crowded field (e.g., it is the first \xrb\ with the detection of X-ray jets, which were originally discovered from these \textit{Chandra} observations; \citealt{corbel02, tomsick03}).  We thus extract sky counts in all seven observations of \jxv\ using a 20-arcsec radius circular aperture centered on a nearby source-free region of the CCD in each image.   Finally, for \jxvi\ we use a 5-arcsec radius circular aperture to extract source counts from obsID 2731 and 10-arcsec radius circular apertures for obsIDs 3400 and 3401 in order to enclose a large number of source counts; we use circular annuli with 20-arcsec inner radius and 30-arcsec inner radius for all three observations of \jxvi.

After extracting spectra for each source, we fit absorbed power laws to each spectrum, and we use the \citet{davis01} pileup model for all three \jxvi\ observations.  These three observations have pileup fractions of $f_{\rm pile} = 0.098$ (obsID 2731), 0.34 (obsID 3400) and 0.32 (obsID 3401), and our best-fit spectral parameters  are similar (within the errors)  to those obtained by \citet{tomsick04}.    We fix the Hydrogen column density in each fit to the values listed in Table~\ref{tab:srcinfo}. The best-fit spectral parameters are listed in Table~\ref{tab:litfits}, which are consistent with previously published values at the 1$\sigma$ level.  We note that our best-fit photon index for the 2000 observation of \viv\ (obsID 97) is significantly softer than the photon index first published by \citet{kong02}.  That observation was likely affected by mild pile-up in the original analysis, as pointed out by \citet{corbel08}.   With  improved reprocessing algorithms  in {\tt CIAO}, the pileup is much less severe in our data reductions.    Our best-fit spectral parameters for this observation  are  consistent with the spectral parameters  presented in \citet{corbel08}, and we additionally find that including the \citet{davis01} pileup model does not substantially improve our fit.

\subsection{Effect of Fixing Column Density in the Spectral Fits}
Since we do not have enough counts to directly fit for \nh\ in all 37 \textit{Chandra} observations, we fix \nh\ to values obtained from the literature (as referenced in Table~\ref{tab:srcinfo}).    These column densities were generally estimated  from higher signal-to-noise X-ray spectra at higher $l_x$, with \nh\ left as a free parameter.   
 Advantages of adopting these \nh\ values include easier comparison to the literature  (since these column densities are often used in other studies), and usually smaller statistical errors on \nh\ than could be obtained from our data.   However, there are also disadvantages.  For one, we must  assume that \nh\ remains relatively constant during the outburst decay.  Most importantly though, fixing \nh\ to predetermined values could  systematically bias  our best-fit photon indices, since using a larger or smaller \nh\ would cause us to overestimate or underestimate $\Gamma$, respectively.  To illustrate the potential magnitude of this effect, we took each system's \textit{Chandra} observation with the highest number of counts (or the second highest number of counts for \maxi\ to avoid pileup), and we refit that spectrum allowing \nh\ to vary as a free parameter.   In Figure~\ref{fig:conf} we show confidence contour maps of $\Gamma$ vs.\ \nh\ for each of these fits.  For comparison, we also show the best-fit values of $\Gamma$ obtained from fixing \nh\  (filled circles; i.e., these circles show the $\Gamma$ values listed in Tables~\ref{tab:fits} and \ref{tab:litfits}).   We omit \jxi\ from Figure~\ref{fig:conf} because it has an extremely small column density, and any systematic effect would be negligible.
 
 The difference between the best-fit photon indices keeping \nh\ fixed vs.\ allowing \nh\ to vary is given in the top-left corner of each panel as $\Delta \Gamma = \Gamma(N_{\rm H,fixed}) - \Gamma(N_{\rm H,free}).$  $\Delta \Gamma$ provides an estimate of the magnitude of any potential systematic bias on $\Gamma$ that could be present for each system (although we note that $\Delta \Gamma$ is approximate, its value is specific to each observation and would not be identical for every observation of the corresponding system).  For about half of the systems, $\Delta \Gamma$ is close to zero and we do not expect a significant systematic bias.  Not surprisingly, the systems with the largest $\Delta \Gamma$ (\hxvii, \maxi, \jxvii, and \gro) also tend to have the fewest number of counts (and therefore the largest uncertainty in \nh).    Thus, we opt to always adopt \nh\ values from the literature for the reasons described earlier, but keeping in mind that this could force a systematic bias at a level shown in Figure~\ref{fig:conf}.
  
  \begin{figure}[t]
\center
\includegraphics[scale=0.48]{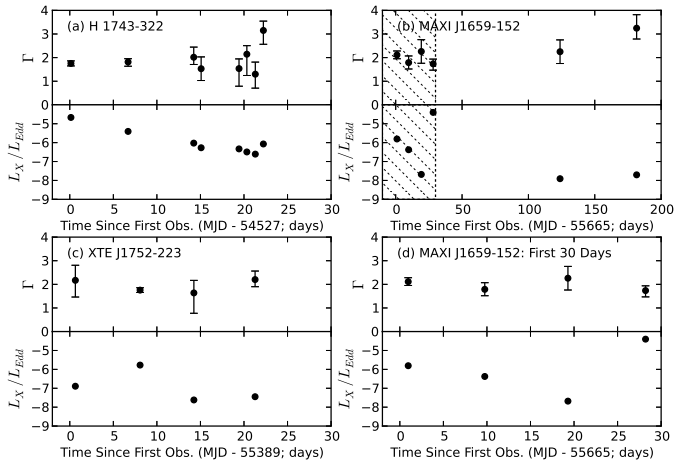}
\caption{Each panel shows photon index vs.\ time (top) and normalized Eddington ratio vs.\ time (bottom) for our three sources with multiple \textit{Chandra} observations at the tail end of an outburst. Each panel's x-axis begins when the first \textit{Chandra} observation was taken.  Panel (b) shows all data for \maxi, while panel (d) shows the same data zoomed in for only the first 30 days of observations (covering the hatched region in panel b).   \maxi\ underwent a small flare during its decay into quiescence.  Otherwise, each source reaches quiescence fairly quickly, and there is no obvious spectral evolution.}
\label{fig:gammaVtime}
\end{figure}

\section{Results}
\label{sec:results}
In Figure~\ref{fig:gammaVtime} we show the temporal evolution of $\Gamma$ and $l_x$ for the  decays of  \hxvii, \maxi, and \jxvii.  The \textit{Chandra} observations cover Eddington ratios between $l_x \sim 10^{-9} - 10^{-4}$.    Each source's X-ray luminosity eventually levels off, indicating that each system has very likely reached quiescence during the observations.       Also, \maxi\ underwent a small flare after it reached quiescence.   There is strikingly little evolution of $\Gamma$ during each decay, even when \maxi\ is flaring.    According to \citet{wu08}, \xrb s attain their hardest X-ray spectra ($\Gamma \sim 1.5$) around $l_x \sim 10^{-2}$.  As \xrb s fade, their spectra soften with decreasing luminosity.    We find that this $\Gamma-l_x$ anti-correlation does not extend to all $l_x < 10^{-4}$, but it eventually saturates to a value near $\Gamma \sim 2$ (Figure~\ref{fig:gammaVlum}).   Such a saturation in quiescence was also suggested by \citet{sobolewska11a}.

\begin{figure}
\center
\includegraphics[scale=0.46]{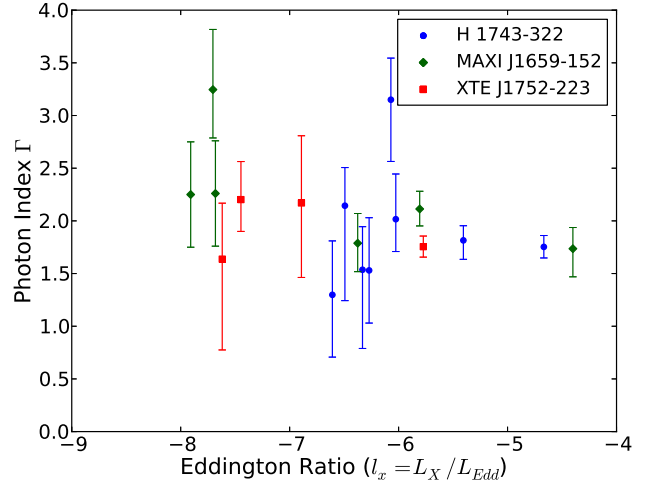}
\caption{Photon Index vs.\  Eddington Ratio for three systems with decays covered by \textit{Chandra.}}
\label{fig:gammaVlum}
\end{figure}

In Figure~\ref{fig:fullsample}a, we add the other 19  \textit{Chandra} observations (see \S\ref{sec:litobs}) to the $\Gamma - l_x$ plane.  For clarity, we also show the same plot  binned by $l_x$ in panel \ref{fig:fullsample}b.   $\Gamma$ does not appear to continue to soften at low-$l_x$ in the full sample either. The 37 observations in Figure~\ref{fig:fullsample} have a Spearman rank correlation coefficient of $\rho = -0.323$ with p=0.051, indicating that there is not a statistically significant anti-correlation between $\Gamma$ and $l_x$ at $l_x < 10^{-4}$.   We cannot definitively identify  the precise luminosity where $\Gamma$ begins to saturate.   However, judging from Figure~\ref{fig:fullsample}, $l_x \sim10^{-5}$ could be a reasonable threshold.   We only have four data points at $l_x > 10^{-5}$, but all four have  similar $\Gamma < 2$ and an average $\langle \Gamma\rangle =  1.70 \pm 0.03$ (although we note one of these data points is from \maxi\ during its mini-flare).  This $\langle \Gamma \rangle$ is  harder than the  average photon index for our remaining 33 observations at $l_x < 10^{-5}$, which have $\langle \Gamma \rangle = 2.08 \pm 0.07$.   Plus, the possibility of an anti-correlation between $\Gamma$ and $l_x$ is even less statistically significant for these 33 observations ($\rho = -0.162$ and $p=0.368$).  Thus, the $\Gamma$-$l_x$ anti-correlation observed for low-hard state \xrb s does not appear to continue into quiescence, and $\Gamma$ seems to plateau by the time $l_x $ reaches $\sim$$10^{-5}$~\ledd.
 
 \begin{figure*}
\center
\includegraphics[scale=0.62]{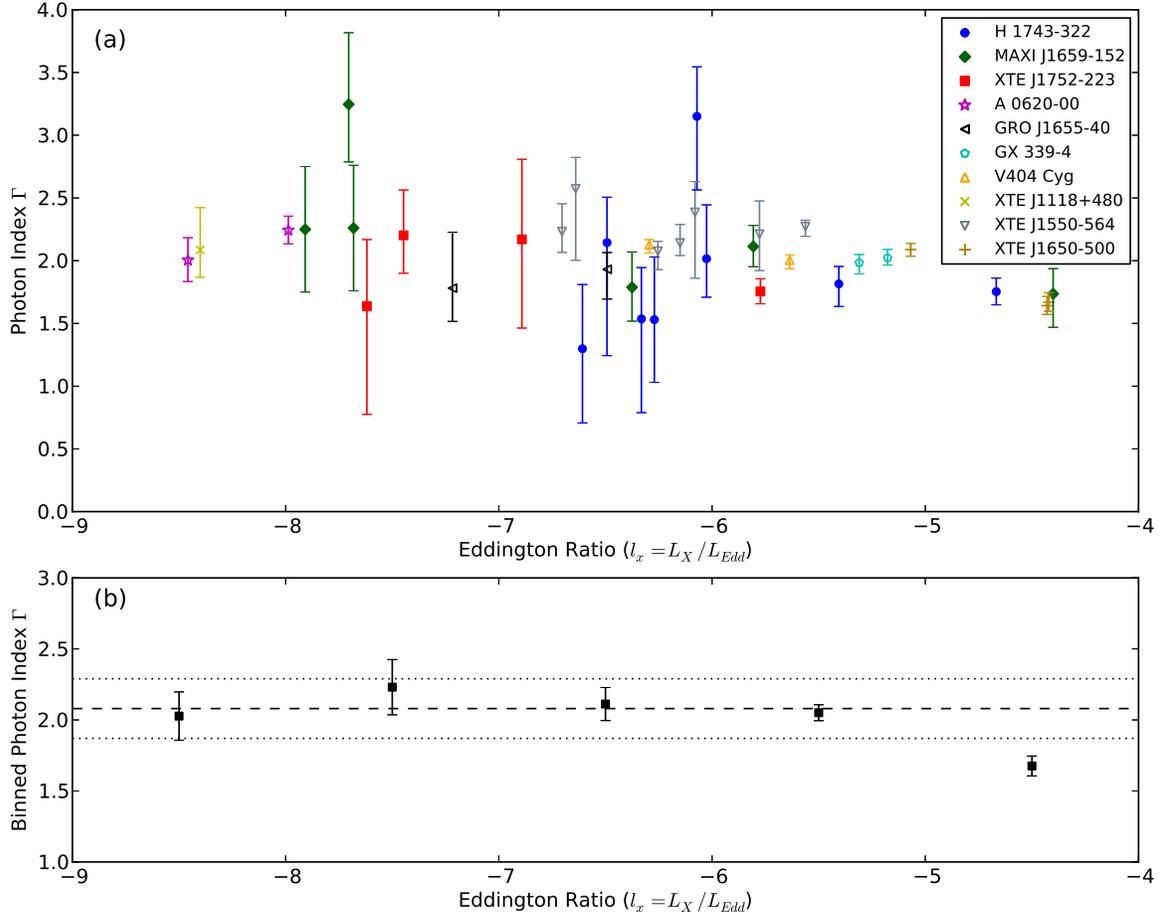}
\caption{(a) Same as Figure~\ref{fig:gammaVlum} but for all ten \xrb\ systems.  (b) Same data  binned by Eddington ratio for clarity.  There is no obvious spectral evolution at $l_x \lesssim 10^{-5}$ on average.  The dashed line in panel (b) marks the average $\left<\Gamma\right>=2.08$ for all observations at $l_x<10^{-5}$, and the dotted lines shows  $\pm$3$\sigma$ deviations.  In the highest luminosity bin, $\Gamma$ is harder by 5.9$\sigma$, indicating that the X-ray spectral softening from the low-hard state heading into quiescence takes place over a relatively small range in $l_x$.}
\label{fig:fullsample}
\end{figure*}

For the 33 observations at $l_x \la 10^{-5}$, we measure a relatively large scatter about $\Gamma$ of $\sigma_{int} = \pm0.39$ (with $\sigma^2_{int}$ defined as the variance about the mean $\langle \Gamma \rangle$ divided by N-1).  This scatter may partly be caused by a low-level of variability in quiescence, but it is more likely that the scatter is primarily due to measurement uncertainty in $\Gamma$ (resulting from the relatively low number of counts in each spectrum).    To illustrate that the scatter could be primarily statistical noise, we randomly draw 33 photon indices from a normal distribution with $\langle \Gamma \rangle = 2.08$ and a standard deviation of $\pm$0.35 (the latter is a typical measurement uncertainty on $\Gamma$ from our spectral fits).  We then measure the scatter of these 33 randomly drawn photon indices about their mean.  We repeat $10^5$ times, and 17\% of our simulated $\Gamma$ distributions have $\left | \sigma_{int} \right | > 0.39$.   We also perform a more detailed test by simulating 10$^3$  X-ray spectra  with $\sim$70 counts (typical of our observations) using an absorbed power-law model with \nh = $5 \times 10^{21}$ cm$^{-2}$ and $\Gamma = 2.08$.  We then fit the simulated spectra keeping \nh\ fixed, but allowing $\Gamma$ to vary.  The distribution for the 10$^4$ best-fit $\Gamma$ has $\sigma_{int} = \pm 0.35$, which is comparable to the observed $\sigma_{int}$ for the 33 observations.  We also repeat the simulations for 10$^3$ spectra with $\sim$30 counts and 10$^4$ spectra with $\sim$200 counts, which show $\sigma_{int} = \pm 0.55$ and $\pm$0.23, respectively. Since the observed scatter is likely dominated by the lowest-count spectra, measurement uncertainty can easily explain the observed scatter.  Thus, to the accuracy of our data, all of the \xrb\ systems have a similar photon index at $l_x \la 10^{-5}$.  

\subsection{Eddington Ratios and Photon Indices at the Lowest  Luminosities}
\label{sec:quiesclum}
The lowest observed X-ray luminosity for each source represents the source's properties deep in quiescence, and the corresponding Eddington ratio and photon index are interesting properties.     Four systems (\gro, \jxi, \jxvi, and \viv) have a single observation that is easily identified as having the lowest observed X-ray luminosity.   Some of our systems, however, have multiple \textit{Chandra} observations at similarly low X-ray luminosities.  Thus, to calculate  the smallest Eddington ratios and corresponding photon indices,  we averaged all data with $L_X < 10^{33.5}$~erg~s$^{-1}$ for \hxvii\ (6 observations), with $L_X < 10^{32}$~erg~s$^{-1}$ for \maxi\ (3 observations), with $L_X < 10^{32}$~erg~s$^{-1}$ for \jxvii\ (2 observations),  and with $L_X < 10^{32.5}$~erg~s$^{-1}$ for \jxv\ (2 observations).     For \avi\ and \gx, we average both observations for each system.    The resulting photon indices and Eddington ratios for all ten systems are shown in Figure~\ref{fig:GammaVQuiescLedd}.    We again see no trend between $\Gamma$ and  Eddington ratio, supporting that (on average)  \xrb s X-ray spectral properties deep in quiescence do not strongly depend on  X-ray luminosity.

\begin{figure}
\center
\includegraphics[scale=0.48]{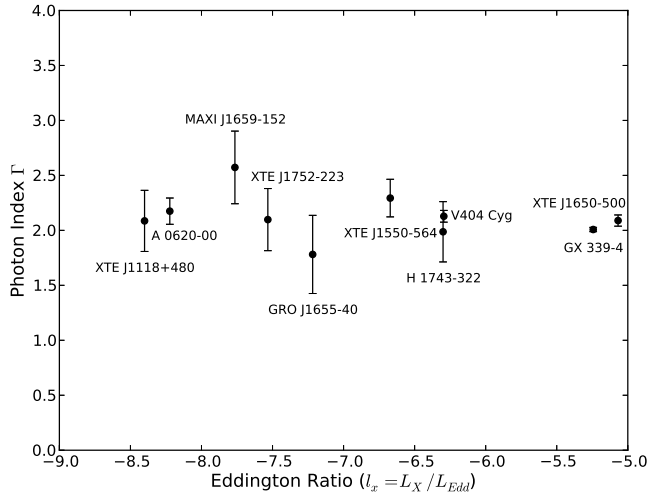}
\caption{Photon Index vs.\ Eddington ratio at the lowest observed X-ray luminosity for each system  (see \S \ref{sec:quiesclum}).}
\label{fig:GammaVQuiescLedd}
\end{figure}

\section{Discussion}
\label{sec:disc}
 
\subsection{X-ray Emission Mechanisms in Quiescence}
\label{sec:adaf}
The fact that we  see X-ray emission from quiescent \xrb s indicates that there is still some amount of  matter falling onto the black hole even at low-$l_x$ (the X-ray emission is unlikely from the corona of the secondary star; see e.g., the discussion in \citealt{narayan08}).    The plateau of $\Gamma$ at low-$l_x$ may represent a spectral signature for when a \xrb\ enters the quiescent state.  Our data suggests that there is either a gradual change in accretion disk structure/geometry between $l_x \sim 10^{-2}$ and at least 10$^{-5}$, or  that the relevant high-energy radiative processes   saturate at some critical luminosity.     There is currently no standardly adopted luminosity threshold in the literature for when a \xrb\ enters quiescence \citep[see][]{remillard06}, and we propose  $l_x < 10^{-5}$ as an observationally-motivated upper luminosity-limit for quiescence.

In the canonical picture for low-hard state \xrb s, some form of a hot, geometrically thick, radiatively inefficient accretion flow (RIAF) is typically invoked for the inner regions of the accretion disk \citep[e.g.,][]{esin97}.  RIAFs (and geometrically thick disks in general) are prone to developing outflows \citep[e.g.][]{rees82, narayan94, narayan95, blandford99, livio99, igumenshchev00, meier01a, hawley02, narayan05},  which is consistent with radio observations that imply that  low-hard state black holes ubiquitously launch compact  jets  \citep[e.g.,][]{corbel00, corbel04, stirling01, miller-jones08, fender09}.   In a RIAF, the amount of X-ray radiation scales non-linearly with accretion rate ($L_X \sim$~\mdot$^q$, typically with $q\sim2$ in the simplest cases).  Thus, only a small amount of the accretion energy is liberated as radiation, thereby resulting in an under-luminous accretion disk compared to a standard cold disk.     The observed (hard) X-rays are  then generated from a combination of inverse Comptonization of lower-energy photons off a hot population of electrons \citep[e.g.][]{malzac01, gilfanov10}, and also a likely contribution from  jet-related  processes \citep[e.g.,][]{falcke95, markoff01, markoff03, markoff05, maitra09}. 

There are many variants of RIAFs in the literature,\footnote{For example, the advection dominated inflow-outflow solution \citep[ADIOS;][]{blandford99} and convection-dominated accretion flows \citep[CDAFs;][]{narayan00, quataert00}.} 
 and the advection dominated accretion flow \citep[ADAF; e.g.,][]{narayan94, abramowicz95} has been particularly successful for modeling the spectra of quiescent \xrb s \citep[e.g.,][]{narayan96, narayan97, hameury97}.   In an ADAF,  the Coulumb coupling between electrons and ions is weak,  resulting in a two-temperature plasma, with hot populations of thermal electrons and ions at characteristic temperatures $T_e$ and $T_i$, respectively  (with $T_e < T_i$).  ADAF models predict an X-ray spectral softening as \xrb s fade into quiescence \citep[e.g.,][]{esin97, qiao13}.  For example, \citet{esin97} show that as $l_x$ decreases,  the optical depth of the hot electrons will decrease and therefore so does the Compton-y parameter.  Thus, as a \xrb\ fades from the low-hard state into quiescence, there is less inverse Compton scattering and fewer hard X-rays are emitted (i.e., the X-ray spectra become softer). The observed inflection point  in the \xrb\ $\Gamma - l_x$ plane at $l_x \sim 10^{-2}$ is likely due to a switch in the source of Comptonized seed photons \citep[e.g.,][]{kalemci05, sobolewska11a, gardner12}.  Thermal seed photons from a cold accretion disk dominate at $l_x \ga 10^{-2}$, while a different source of seed photons dominate at lower $l_x$ (e.g., cyclo-synchrotron from the hot accretion flow or synchrotron self-Compton are two possibilities).

 \citet{esin97} predict that the softening of the X-ray spectrum (from 1-10~keV) peaks at $\Gamma \sim 2.2$.   Then in quiescence, inverse Compton scattering becomes so weak  that bremsstrahlung radiation will start to dominate the hard X-rays, which will slightly re-harden the X-ray spectrum  to $\Gamma \sim 1.7$ (which is  harder than our X-ray observations on average, but not inconsistent given our individual error bars).  \citet{yuan05} point out that at low accretion rates ($l_x \la 10^{-5} - 10^{-6}$) the jet should start to dominate in the X-ray, instead of bremsstrahlung.   They consider an ADAF coupled to a compact steady-state jet, and they include  radiative losses from synchrotron cooling in their jet.     We expect such synchrotron cooled radiation to follow a power-law with $\Gamma \sim 2$, also consistent with our observations.    Since the jet spectrum is a power-law, we would naturally expect $\Gamma$ to plateau in quiescence.  We note that \citet{gardner12} also include a jet in their RIAF model for \xrb s, but they do not find   that the jet ever dominates quiescent X-ray spectra.  However, \citet{gardner12}  do not include synchrotron cooling losses in their model.    

The shape of the spectrum is a major difference  between jet-dominated and RIAF-dominated emission models.  Jets should produce power-law X-ray spectra in quiescence, while ADAF  X-ray spectra  (and spectra from most types of RIAFs in general) should be curved since their hard X-rays are inverse Compton scattered off of a thermal distribution of electrons.\footnote{However, a curved X-ray spectrum is not expected from RIAF models with non-thermal electron distributions.} 
   An interesting prediction of RIAF models is that, since electrons have higher $T_e$ and lower optical depth in quiescence, one expects to see multiple Compton peaks (opposed to the low-hard state where  multiple Compton peaks merge together into one broad hump; \citealt{mcclintock03}).  Thus, the amount of curvature expected in quiescent X-ray spectra depends on which scattering order falls into the  X-ray band.   Unfortunately, there is not any existing X-ray spectrum that can sufficiently constrain the X-ray spectral shape well enough to differentiate between jet and RIAF X-ray origins.
  
Finally, we note that there is  evidence, at least for more luminous low-hard state \xrb s ($l_x \ga 10^{-3}$),  that a standard cold accretion disk could always extend close to the innermost stable circular  orbit  \citep[ISCO; e.g.,][]{miller06a, miller12a, malzac07, wilkinson09, reynolds10, reis10, uttley11}.  However, in quiescence the accretion flow is so underluminous that it seems unlikely that  a cold accretion disk could be present close to the ISCO.  We do not see an obvious soft X-ray component that would require the addition of a cold disk to our spectral models (although it is not clear if such a component could be easily seen in our low-count spectra).  Regardless, in quiescence, the standard picture invoking a RIAF in the inner regions  (and a standard disk in the outer regions) with the addition of a possible jet component seems plausible.  However, we stress that there are other possibilities  \citep[also see, e.g.,][]{merloni02, xie12}.

\subsection{On the Transition from the Low-hard State into Quiescence}
\label{sec:rxcorr}
 Many low-hard state \xrb s  show a relatively tight non-linear correlation between radio and X-ray luminosity \citep[][although see \citealt{gallo12}]{corbel03, gallo03a}, which may imply a coupling between the accretion flow and the radio  jet.    The observed radio emission is optically thick (i.e., self-absorbed) synchrotron radiation from a compact steady jet.   In order to explain the observed slope of the luminosity correlation, the X-ray radiation must be inefficient with $L_x \sim$~\mdot$^2$ \citep{gallo03a, markoff03, markoff05}.  Such inefficient X-ray emission is consistent with either a RIAF \citep{merloni03} or with optically thin jet synchrotron emission \citep{markoff03, falcke04, plotkin12a}.    The  radio/X-ray luminosity correlation has also been extended to include the optical and infrared (OIR) wavebands \citep{homan05a, russell06, coriat09}.   OIR/X-ray correlations also imply a coupling between the disk and jet (with the OIR dominated by non-thermal jet radiation and/or reprocessed disk emission).
 
The compact steady jet has three main components.  At the lowest frequencies is the optically thick part showing a flat or inverted spectrum, as is typically observed in the radio.  At higher frequencies the jet becomes optically thin (usually around the infrared for a $\sim$10~$M_\odot$ black hole), typically showing $\Gamma$ between 1.5--1.7.  Finally, at the highest frequencies the jet  becomes synchrotron cooled (i.e., radiation losses become so large that the emitting particles lose kinetic energy), where we expect $\Gamma$ to steepen (i.e., increase) by $\sim$0.5.    Considering this expected shape for the jet's broadband spectrum and the   $\left<\Gamma\right>=2.08$ observed from our sample, if the jet  dominates in the  X-ray waveband in quiescence, then we expect that jet emission should  already be synchrotron cooled.   In this case, the observed X-ray luminosity will scale \textit{linearly} with \mdot\ \citep[e.g.,][]{heinz04, yuan05}, instead of approximately \textit{quadratically} as with either RIAF or optically thin jet synchrotron.  Then, the slope of the radio/X-ray  luminosity correlation should be different in quiescence, as suggested by \citet{yuan05} who predict that the  correlation steepens to $L_R \sim L_X^{1.23}$ below $l_x \la 10^{-5} - 10^{-6}$ (the exact slope and transition luminosity depend on specific model parameters).   One would similarly expect the OIR/X-ray correlation to steepen as well.  

Exactly how the transition to synchrotron cooled dominated X-rays occurs depends on the frequency of the jet cooling break $\nu_{cool}$ (i.e,  the frequency where the jet transitions from being optically thin to synchrotron cooled).  If $\nu_{cool}$ is always below the X-ray band, then RIAF emission could dominate in the low-hard state with $L_X \sim$ \mdot$^2$) and the synchrotron cooled part of jet would dominate in quiescence (with $L_X \sim$ \mdot).    However, it is unclear if $\nu_{cool}$ always falls at such low frequencies, and therefore if such a transition from RIAF to synchrotron cooled  X-ray emission actually occurs in nature.    For example, at least some  low-hard state \xrb s can have optically thin synchrotron emission extending into the X-ray waveband \citep[e.g.,][]{markoff01, russell10}.    Also, in several hard-state systems a high-energy break that could be associated with synchrotron cooling has been observed  at a few tens of keV (i.e., above the \textit{Chandra} X-ray band, although that break could instead be associated with cooling from Comptonization; see \citealt{peer12}).
 
 Given the above uncertainties on the location of $\nu_{cool}$, another plausible scenario to explain the observed spectral softening is that $\nu_{cool}$ is above X-ray energies in most low-hard state \xrb s.   Then, as \xrb s fade into quiescence, $\nu_{cool}$ shifts through the X-ray band.   In this case,  X-rays could always be jet dominated, but jet X-rays would be optically thin synchrotron in the low-hard state and synchrotron cooled in quiescence.   \citet{russell13} isolate the frequency where jets transition from optically thick to optically thin  for 12 \xrb s at various $l_x$ (this optically thin transition happens at lower frequencies than the cooling break, namely $10^{12} - 10^{14}$~Hz).  The optically thin jet break is related to the location along the jet where particles are accelerated into a non-thermal distribution \citep[see e.g.,][]{polko13}.  From their sample, \citet{russell13}  speculate that the higher-energy cooling break $\nu_{cool}$ could shift from hard X-ray energies at $l_x \sim 10^{-3}$ to the ultraviolet at $l_x \sim 10^{-5}$.     However, more observations are needed to determine  how $\nu_{cool}$ evolves as \xrb s fade into quiescence,  especially since factors other than \mdot\ are almost certainly important for determining the evolution of $\nu_{cool}$ and  jet properties in general. For example, if the magnetic field $B$ at the location along the jet where synchrotron radiative losses becomes important depends \textit{only} on \mdot, then one would expect the exact opposite evolution where $\nu_{cool}$ should instead increase with decreasing \mdot.\footnote{The typical Lorentz factor of a synchrotron cooled electron is $\gamma_{cool} \sim \left( B^2 t_{cool} \right)^{-1}$,  where $t_{cool}$ is the dynamical cooling timescale  \citep[e.g.][]{rybicki79, heinz04}.  Thus, $\nu_{cool} \sim B \gamma_{cool}^2 \sim$~\mdot$^{-3/2}$, if $B^2\sim$~\mdot\ (as expected for a mechanically cooled flow; \citealt{heinz03}).}  
  Also, general relativistic magnetohydrodynamic simulations show that beyond \mdot, the geometry of the magnetic field  threading the disc  close the the black hole can influence whether a strong collimated  jet is even launched in the first place \citep{mckinney09, dibi12}.  There is also the possibility of inverse Compton scattering off of jet particles contributing to the observed X-ray emission (either from synchrotron or external seed photons).  Thus, the expected evolution of $\nu_{cool}$ as \xrb s fade into quiescence is unclear, and more detailed studies are needed.

Further complicating matters is that the radio/X-ray luminosity correlation for low-hard state \xrb s is not as universal as once thought.  Recently, there have been discoveries of several outliers to the ``standard'' correlation, such that at a given X-ray luminosity these outliers are fainter in the radio than expected \citep[e.g.,][]{corbel04, brocksopp05, cadolle-bel07, jonker10, soleri10, ratti12}.  There is now statistical evidence for two tracks within the radio/X-ray luminosity correlation, with the ``standard''  track following $L_r \sim L_X^{0.63}$ and the ``radio-faint'' track following a steeper slope $L_r \sim L_X^{0.98}$ \citep{gallo12}.     It is interesting that the only two systems with firm simultaneous radio and X-ray detections in quiescence --- \avi\ at $l_x \sim 10^{-8.5}$ \citep{gallo06} and \viv\ at $l_x \sim 10^{-6.6}$ \citep{gallo05, miller-jones08} --- fall  along the ``standard track'' extrapolated to low-$l_x$.    All other efforts to detect radio emission in quiescence have only yielded upper limits so far, and we have not yet seen any conclusive evidence for radio-faint \xrb s at $l_x \la 10^{-5}$ \citep[see][and references therein.  Also see Figure~9 of \citealt{corbel13} for a recent radio/X-ray correlation including radio-faint systems]{miller-jones11}.   Some low-hard state \xrb\ systems (including \hxvii, \maxi, and \jxvii\ in our sample)  have  even been observed to switch from the ``radio-faint'' track  to the ``standard track'' by the time they reach $l_x \sim 10^{-5}$ \citep[e.g.,][]{jonker10, coriat11, jonker12, ratti12}.  While perhaps coincidental, it is potentially interesting that the ``radio-faint'' track seems to end around the same $l_x$ where we find $\Gamma$ saturates in our \xrb\ sample.  We speculate that this similarity in $l_x$ could  imply that the same type of high-energy emission processes tend to dominate in most quiescent systems, while there is potentially more variety in the low-hard state \citep[although further work is needed, see e.g.,][]{calvelo10}.

 \subsection{Orbital Parameters}
 There is a well-known relationship between the lowest observed X-ray luminosity deep in quiescence and orbital period, where \xrb\ systems with the faintest X-ray luminosities have the shortest  periods \citep[e.g.,][]{lasota98, menou99, garcia01}.\footnote{A trend between  X-ray luminosity and orbital period is also observed for neutron star X-ray binaries, although neutron star systems are more luminous than \xrb s at comparable orbital periods.} 
  The X-ray luminosity -- period relationship indicates that  for systems with shorter periods, the secondary star transfers less mass per unit time onto the accretion disk (assuming that a similar fraction of transferred mass eventually accretes onto the black hole in all systems).  The mass transfer rate is expected to depend on orbital period, since systems with similar orbits are likely to have similar types of secondary stars (with similar mass loss rates).  Also, different mechanisms may drive the mass transfer at different orbital periods.  For example, see \S 3 of \citealt{menou99} (and references therein) for a discussion on angular momentum losses from gravitational radiation (and perhaps magnetic breaking) at short orbital periods compared to mass loss driven by the secondary's nuclear evolution at longer periods.

Our ten systems are consistent with the  X-ray luminosity -- period relationship (although see \citealt{jonker12} regarding \maxi\ perhaps being brighter than expected).   One might thus expect that the orbit could also affect the shape of quiescent \xrb\ spectra, as proposed by \citet{corbel06} who found that three long-period systems (\gro, \viv, and V4641~Sgr) show harder X-ray spectra deep in quiescence compared to shorter-period systems.     However,  we do not see any trend between our ten systems' $\Gamma$ at their faintest $L_X$ and their  orbital parameters (specifically period and inclination; see Figure~\ref{fig:orbit}).  The lack of a dependence of $\Gamma$ on orbital parameters is consistent with the conclusion of \citet{corbel08}, who re-examined the quiescent properties of \viv, which is the source that  (statistically) dominated the earlier \citet{corbel06} study.     Although we find that \gro\ may be slightly harder than other quiescent \xrb\ systems, it is not harder at a statistically significant level.\footnote{\citet{corbel06} include a harder $\Gamma=1.30^{+0.34}_{-0.41}$ data point for \gro\ in their study, which they took from \citet{hameury03}.  This  harder $\Gamma$ may be due to the choice of column density, as it was obtained by fixing \nh$=6.7\times10^{21}$~cm$^{-2}$ when fitting an \textit{XMM-Newton} spectrum.  \citet{hameury03}, however,  also obtain a steeper $\Gamma=1.54^{+1.02}_{-0.72}$ when allowing \nh\ to vary as a free parameter.  In this work, we adopt  a slightly larger column density of \nh$=8.1\times10^{-21}$ cm$^{-2}$ when fitting our \textit{Chandra} spectra, which yields a  steeper $\Gamma$ (and we show in Figure~1 that we might still be systematically underestimating both \nh\  and $\Gamma$, if anything). }  
 The final source with a hard quiescent X-ray spectrum in \citet{corbel06}, V4641~Sgr, really does seem to stay hard in quiescence.  However, a power-law appears to be a poor fit to \textit{Chandra} observations of this source at low-$l_x$, and its hard spectrum is unlikely  driven by orbital period.  We will discuss this interesting source in more detail in a future paper (Gallo, Plotkin, \& Jonker in prep).
 
 \begin{figure}
\center
\includegraphics[width=3.8in, height=4.5in]{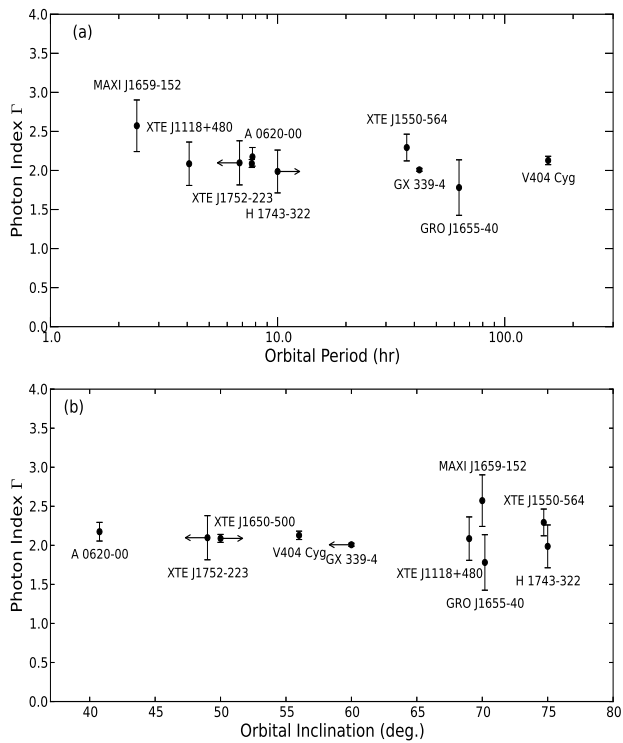}
\caption{There is no dependence of $\Gamma$  on the binary system orbital parameters.  (a)  $\Gamma$ at the lowest observed X-ray luminosity vs.\ orbital period.  The label for \jxvi\ at P=7.7 hr is omitted for clarity.   (b)  $\Gamma$ vs.\ inclination.}
\label{fig:orbit}
\end{figure}

 In summary, the orbital period and rate of mass transfer can affect the total amount of emitted X-rays.  However,  by the time matter reaches the inner regions of the accretion flow, it is apparently the properties/geometry of the accretion disk (and potential connections to any outflow) that most strongly control how the X-rays are produced.   In other words, the accretion disk feeding mechanism may determine the X-ray ``normalization'', but it does not strongly affect the emission mechanism deep in quiescence.    A similar argument would suggest that black hole spin is also unlikely very important for controlling X-ray spectral properties in quiescence.

 \subsection{Comparisons to Supermassive Black Holes}
 Changes in \mdot\ not only affect the accretion states of \xrb s, but also may be partly responsible for different subclasses of AGN. For example, the ``standard track'' of the \xrb\ radio/X-ray luminosity correlation can be extended to include supermassive black holes by incorporating an additional mass normalization term \citep[i.e., the fundamental plane of black hole activity;][]{merloni03, falcke04}.   Supermassive analogs to low-hard state black holes include low-luminosity AGN (LLAGN) and low-luminosity radio galaxies (i.e., FR~Is and BL~Lac objects) \citep[e.g.,][]{falcke04, plotkin12a}.    These supermassive analogs do not usually show strong ``big blue bumps'', indicating significantly weaker disk emission than expected from a standard cold accretion disk \citep{ho08}.   Like low-hard state \xrb s, these AGN also tend to show compact radio emission from a jet \citep[e.g.,][]{nagar05, sikora07}, and their X-ray emission can  usually be  modeled with a hot inner accretion flow like a RIAF (sometimes  with a potential jet contribution; e.g., \citealt{quataert99, nemmen06, ptak04, yu11}).    ``Low-hard state'' AGN also tend to show weak broad emission lines and dusty tori, which can also be explained (at least qualitatively) by invoking a RIAF \citep[e.g.,][]{nicastro00, ghisellini01, elitzur06,  plotkin12}.

 In addition to the above similarities, LLAGN also  show an anti-correlation between $l_x$ and $\Gamma$ at low-$l_x$ that is similar to the observed anti-correlation for low-hard state \xrb s \citep[][although also see \citealt{winter09}]{constantin09, gu09, younes11, gultekin12}.\footnote{Observed X-ray luminosities represent different fractions of the total bolometric luminosity for AGN and for \xrb s, which  studies focusing on AGN take into account.   Here, we continue to use the $l_x$ notation even for AGN simply for convenience (instead of $L_{bol}/L_{Edd}$).}  
 We would thus  expect supermassive black holes' hard X-ray photon indices to  also plateau when they are in an analogous quiescent state.   LLAGN with $10^{-8} \la l_x \la 10^{-4}$ generally show $1.5\la \Gamma \la 2.5$ \citep[e.g.,][]{soria06, zhang09, younes11}, consistent with our expectations.  However, a similar type of plateau at very low-$l_x$ has yet to be seen, which could be due to a combination of large error bars and not enough source statistics.

\section{Summary}
\label{sec:conc}

We followed the X-ray spectral evolution of three \xrb\ systems (\hxvii, \maxi, and \jxvii) with \textit{Chandra} during the final parts of their outburst decays.  We focus on $l_x \la 10^{-4}$ here because \xrb\ X-ray spectral properties are relatively unconstrained in this luminosity regime.  While these three systems' X-ray spectra are softer than typical low-hard state \xrb s, there is little to no spectral evolution for these three systems at $l_x \la 10^{-4}$.   We  compare these systems to all other \xrb\ systems with available \textit{Chandra} spectral coverage at $l_x \la 10^{-4}$, adding another seven systems to our sample.  We find that the anti-correlation between $\Gamma$ and $l_x$ in the low-hard state  at $l_x \la 10^{-2}$ does not extend all the way into quiescence.  Rather, by the time a \xrb\ reaches $l_x \sim 10^{-5}$, its X-ray spectrum saturates to $\langle \Gamma \rangle \sim 2.08 \pm 0.07$ on average (with the observed scatter about $\langle \Gamma \rangle$ likely dominated by measurement error).  Our  \xrb\ X-ray spectra do not appear to depend on the binary systems' orbital parameters.  Therefore, it is probably not the feeding mechanism but rather  the properties of the accretion flow itself that most strongly determine how X-rays are produced.   The similar $\Gamma$ observed for each quiescent \xrb\ system, and the lack of any ``radio-faint'' \xrb s observed at $l_x \la 10^{-5}$ so far,  might indicate that there is less variety in X-ray processes in quiescence compared to the low-hard state.  However, more studies are needed.   Based on the X-ray spectral properties of our low-$l_x$ \xrb\ sample,  $l_x \sim 10^{-5}$ seems to be an observationally-motivated  luminosity threshold for when a \xrb\ enters the quiescent state.

In the future, higher signal-to-noise X-ray spectra than currently available will be critical in order to distinguish between RIAF vs.\ jet.\ dominated X-rays (e.g., most RIAF models predict curved X-ray spectra, while jets predict pure power-laws).   X-ray polarimetry would also constrain the  geometry of the emission regions \citep{laurent11}.   In order to determine if \xrb s switch from RIAF to jet dominated X-ray emission in quiescence, more radio constraints are needed to measure  the slope of \xrb\ radio/X-ray correlations in quiescence \citep[e.g.,][]{yuan05}, and also to continue to search for ``radio-faint'' \xrb s at $l_x \la 10^{-5}$ \citep{miller-jones11}.    More observational constraints on the location of the jet synchrotron cooling break (which might be possible with the Nuclear Spectroscopic Telescope Array [nuSTAR]) would also be helpful. 

\textit{Is quiescence an extension of the low-hard state or rather a distinct spectral state?}  Although we argue that the plateau in $\Gamma$ at $l_x \la 10^{-5}$ is an observational signature that a \xrb\ has entered quiescence, we cannot easily envision a  scenario where the observed plateau in $\Gamma$ indicates a very significant change in the accretion flow.  That is, there are subtle differences between the observed emission from quiescent and low-hard state \xrb s, and there is thus some value to thinking of quiescence as its own spectral state.  However,  quiescence does not appear to represent a distinct spectral state  to the same degree as the soft-to-hard or the hard-to-soft transitions at higher $l_x$.      As \xrb s fade, either a jet's cooling break shifting through the X-ray band or inverse Compton processes becoming less dominate (owing to lower optical depths) can potentially explain their X-ray emission.  Regardless of the right answer though, the basic accretion disk/outflow structure and geometry is likely the same in quiescence as in the low-hard state.

\acknowledgments
We thank the anonymous referee for helpful comments that improved this manuscript.  Support for this work was provided by the National Aeronautics and Space Administration through Chandra Award Number GO1-12049A issued by the Chandra X-ray Observatory Center, which is operated by the Smithsonian Astrophysical Observatory for and on behalf of the National Aeronautics Space Administration under contract NAS8-03060.  P.G.J. acknowledges support from the Netherlands Organisation for Scientific Research.   This research has made use of data obtained from the Chandra Data Archive and the Chandra Source Catalog, and software provided by the Chandra X-ray Center (CXC) in the application packages CIAO, ChIPS, and Sherpa.




\begin{table*}
\caption{Source Parameters}
\label{tab:srcinfo}
\scriptsize
\begin{tabular}{l c c c c c c c }
\hline \hline
	Source Name       &  
	RA   &  
	Dec.                        & 
	\nh\ $\times 10^{22}$ &  
	Mass\tablenotemark{a}                     & 
	Distance                & 
	P$_{\rm orb}$       & 
	Inclination             \\ 
	                                & 
	(J2000)                            & 
	 (J2000)                               & 
	 (cm$^{-2}$)          & 
	 (M$_{\sun}$)          & 
	 (kpc)                     & 
	 (hr)                        & 
	 (deg)                \\ 
\hline	 
\multicolumn{8}{l}{\textit{Sources from Jonker et al.\ \textit{Chandra} TOO programs}} \\
H 1743-322            &  $17^{\rm h}46^{\rm m}15^{\rm s}.61$   &   $-32^\circ14\arcmin00.6\arcsec$(1)      &      2.3(2,3)   &    ...      &   $8.5\pm0.8$(4)    &   $\ga$10(3)  & $75 \pm 3$(4) \\
MAXI J1659-152    & $16^{\rm h}59^{\rm m}01^{\rm s}.68$  &   $-15^\circ15\arcmin28.7\arcsec$(5)  &    0.23(6,7)   &  ...      &   $6\pm2$(7)        &   2.4(6,8) &  $60-75$(6) \\
XTE J1752-223      &  $17^{\rm h}52^{\rm m}15^{\rm s}.09$   &   $-22^\circ20\arcmin32.4\arcsec$(9)     &       0.5(10,11)     &    $9.6\pm0.9$(12) &  3.5--8(11)       &  $\la$$6.8$(11)  & $<$49(9,13) \\

\multicolumn{7}{l}{\textit{Other systems with  \textit{Chandra} X-ray spectra}} \\
A 0620-00  &    $06^{\rm h}22^{\rm m}44^{\rm s}.54$     &   $-00^\circ20\arcmin44.4\arcsec$(14)        &       0.194(15) &    $11.0 \pm 1.9$(16)   &  $1.16 \pm 0.11$(16)   &  7.75(17) & $40.75 \pm 3$(16) \\
GRO J1655-40   &    $16^{\rm h}54^{\rm m}00^{\rm s}.20$     &    $-39^\circ50\arcmin43.6\arcsec$(18)        &       0.859(15) &    $6.3 \pm 0.6$(19)       &  $3.2 \pm 0.2$(20,21)    &   62.92(20,21) & $70.2 \pm 1.9$(19) \\
GX 339-4             &   $17^{\rm h}02^{\rm m}49^{\rm s}.50$   & $-48^\circ47\arcmin23.0\arcsec$(22)  &    0.6(23)  &    $5.8\pm0.5$(24) &  $8\pm4$(21,25) &  42.1(24)  & $<$60(26)  \\
V404 Cyg\tablenotemark{b}       &    $20^{\rm h}24^{\rm m}03^{\rm s}.82$     &     $\ \ 33^\circ52\arcmin01.9\arcsec$(27,28)        &        0.81(29,30)    &    $12 \pm 2(31) $  &  $2.39 \pm 0.14$(28) &    155.28(32) & $56 \pm 4$(31) \\
XTE J1118+480\tablenotemark{c}  &    $11^{\rm h}18^{\rm m}10^{\rm s}.85$    &     $\ \ 48^\circ02\arcmin12.9\arcsec$(33)		    &         0.012(34)   &   $6.9-8.2$(35)   &  $1.72 \pm 0.10$(36)    &    4.08(37)  &  $68-79$(35) \\
XTE J1550-564  &    $15^{\rm h}50^{\rm m}58^{\rm s}.78$     &     $-56^\circ28\arcmin35.0\arcsec$(38)			    &        0.9(39,40)        &   $9.10 \pm $ 0.61(41)  &  $4.38^{+0.58}_{-0.41}$(41)   &  37.03(41) & $74.7 \pm 3.8$(47) \\
XTE~J1650-500 &    $16^{\rm h}50^{\rm m}00^{\rm s}.98$ &     $-49^\circ57\arcmin43.6\arcsec$(42)           &    0.67(42)        &   $<$7.5(43)    &   $2.6 \pm 0.7(44) $     &   7.7(43)  & $>50 \pm 3$(43) \\
\hline
\end{tabular}
\\ 
\tablenotetext{1}{We assume $M = 10 M_{\odot}$ for \hxvii\ and \maxi.} 
\tablenotetext{2}{\viv\ has a proper motion of $\mu_\alpha cos \delta =  -5.04 \pm 0.02$ milli-arcsec yr$^{-1}$ and $\mu_\delta = -7.64\pm 0.02$ milli-arcsec yr$^{-1}$.  The coordinates given in columns 2 and 3 are referenced to MJD 54322 \citep{miller-jones09}.}
\tablenotetext{3}{\jxi\ has a proper motion of $\mu_\alpha  = -16.8 \pm 1.6$ milli-arcsec yr$^{-1}$ and $\mu_\delta = -7.4 \pm 1.6$ milli-arcsec yr$^{-1}$ \citep{mirabel01}.  The coordinates given in columns 2 and 3 are referenced to MJD 51635 \citep{garcia00}.}

References: (1) \citet{steeghs03}, (2) \citet{miller06}, (3) \citet{jonker10}, (4) \citet{steiner12}, (5) \citet{paragi10}, (6) \citet{kennea11}, (7) \citet{jonker12}, (8) \citet{kuulkers10}, (9) \citet{miller-jones11a}, (10) \citet{curran11}, (11) \citet{ratti12}, (12) \citet{shaposhnikov10}, (13) \citet{reis11}, (14) \citet{gallo06}, (15) \citet{kong02}, (16) \citet{gelino01}, (17) \citet{mcclintock86}, (18) \citet{hjellming94}, (19) \citet{greene01}, (20) \citet{jonker04a} (21) \citet{dunn10}, (22) \citet{reynolds11a}, (23) \citet{corbel13}, (24) \citet{hynes03}, (25) \citet{zdziarski04}, (26) \citet{cowley02}, (27) \citet{miller-jones08} (28) \citet{miller-jones09}, (29) \citet{bradley07}, (30) \citet{corbel08}, (31) \citet{shahbaz94}, (32) \citet{casares92}, (33) \citet{garcia00}, (34) \citet{mcclintock03}, (35) \citet{khargharia13}, (36) \citet{gelino06}, (37) \citet{cook00}, (38) \citet{jain99}, (39) \citet{tomsick01}, (40) \citet{corbel06}, (41) \citet{orosz11}, (42) \citet{tomsick04}, (43) \citet{orosz04}, (44) \citet{homan06}

\end{table*}
 \begin{table*}[htbp]
 \center
\caption{Log of  \textit{Chandra} X-ray Observations Tracking Individual Outburst Decays}
\label{tab:obslog}
\begin{tabular}{l c c c c c c}
 \hline \hline
                Obs.     		&  
                Obs.            &  
                MJD         		&  
                Time on             &   
                Net count rate                       &  
                Net source		&  
                Background 	\\ 
	      ID	 			&  
	      date			& 
	      (d; UTC)			& 
	      source (ks)		& 
	      0.5--7 keV (counts s$^{-1}$)	 & 
	      counts             &  
	      counts                \\ 
	      
\hline

\multicolumn{2}{l}{\textit{H 1743-322}} &    &     &      &   &   \\ 
8987   & 2008 Mar 02  &  54527.13111 &    6.5 & $(9.92 \pm 0.41) \times 10^{-2}$    & 643.0       & 1.0      \\ 
8988   & 2008 Mar 08  &  54533.69200 &   13.7 & $(1.80 \pm 0.12) \times 10^{-2}$    & 247.8      & 1.2      \\ 
8989   & 2008 Mar 16  &  54541.23027 &   20.5 & $(3.56 \pm 0.47) \times 10^{-3}$    & 73.2        & 0.8      \\ 
9833   & 2008 Mar 17  &  54542.08033 &   11.0 & $(2.04 \pm 0.53) \times 10^{-3}$    & 22.5        & 0.5      \\ 
9838   & 2008 Mar 21  &  54546.42945 &   23.8 & $(1.74 \pm 0.32) \times 10^{-3}$    & 41.4        & 0.6      \\ 
8990   & 2008 Mar 22  &  54547.32918 &   21.2 & $(1.29 \pm 0.30) \times 10^{-3}$    & 27.4        & 0.6      \\ 
9839   & 2008 Mar 23  &  54548.30002 &   28.7 & $(1.19 \pm 0.24) \times 10^{-3}$    & 34.1       & 0.9      \\ 
9837   & 2008 Mar 24  &  54549.22579 &   20.6 & $(1.58 \pm 0.33) \times 10^{-3}$    & 32.5       & 0.5      \\ 
\multicolumn{2}{l}{\textit{MAXI J1659-152}} &    &     &      &   &   \\ 
12438  & 2011 Apr 14  &  55665.96202 &    6.4 & $(3.73 \pm 0.26) \times 10^{-2}$    & 236.8      & 0.2      \\ 
12439  & 2011 Apr 23  &  55674.74944 &    9.1 & $(8.67 \pm 1.09) \times 10^{-3}$    & 78.7        & 0.3      \\ 
12440  & 2011 May 03  &  55684.29844 &   13.6 & $(5.61 \pm 2.91) \times 10^{-4}$    & 7.6        & 0.4      \\ 
12441  & 2011 May 12  &  55693.21054 &   18.1 & $(6.74 \pm 0.06) \times 10^{-1}$    & 12226.4    & 6.6      \\ 
12442  & 2011 Aug 15  &  55788.83283 &   30.8 & $(3.31 \pm 1.44) \times 10^{-4}$    & 10.2      & 0.8      \\ 
12443  & 2011 Oct 12  &  55846.53179 &   90.7 & $(4.34 \pm 0.82) \times 10^{-4}$    & 39.3       & 1.7      \\ 
\multicolumn{2}{l}{\textit{XTE J1752-223}} &    &     &      &   &   \\ 
11053  & 2010 Jul 12  &  55389.63924 &    6.4 & $(3.43 \pm 0.91) \times 10^{-3}$    & 21.8      & 0.2      \\ 
12310  & 2010 Jul 20  &  55397.07034 &   13.6 & $(3.86 \pm 0.18) \times 10^{-2}$    & 525.5      & 1.5      \\ 
11055  & 2010 Jul 26  &  55403.24698 &   31.4 & $(6.16 \pm 1.77) \times 10^{-4}$    & 19.4       & 0.6      \\ 
11056  & 2010 Aug 02  &  55410.27430 &   88.9 & $(8.04 \pm 1.08) \times 10^{-4}$    & 71.5      & 1.5      \\ 
\hline
\end{tabular}
\\[0.1cm]
Notes: Data were originally published in \citet{jonker10}, \citet{jonker12}, and \citet{ratti12} for \hxvii, \maxi, and \jxvii, respectively.

\end{table*}

\renewcommand\arraystretch{1.1}

\begin{table*}[htbp]
\center
\caption{Best-fit Spectral Parameters for  \textit{Chandra} Observations Tracking  Outburst Decays}
\label{tab:fits}
\renewcommand{\thefootnote}{\thempfootnote}
\centering
\begin{tabular}{l c c  c c c}
 \hline \hline
                Obs.      		&  
                MJD            &  
                $\Gamma$\tablenotemark{a}             &   
                Unabs.\ 0.5--10 keV flux\tablenotemark{a,b}                      &  
                $l_x$\tablenotemark{c} & 
                Goodness\tablenotemark{d}\\ 
	       ID	 			&  
	       (d; UTC)		& 
	      				& 
	      (erg cm$^{-2}$ s$^{-1}$)	 & 
	      & 
	                            \\ 
\hline	                          
\multicolumn{2}{l}{\textit{H 1743-322}}  &    &     &          &     \\ 
8987                &  54527.13111   & $1.75 \pm 0.15$           & $(4.0 \pm 0.7) \times 10^{-12}$         & $2 \times 10^{-5}$          &  0.31         \\ 
8988                 &  54533.69200  & $1.81 \pm 0.23$           & $(7.4 \pm 1.9) \times 10^{-13}$         & $4 \times 10^{-6}$          & 0.43        \\ 
8989                 &  54541.23027  & $2.02 \pm 0.53$           & $(1.8 \pm 1.0) \times 10^{-13}$         & $9 \times 10^{-7}$    & 0.55        \\ 
9833\tablenotemark{e} &  54542.08033   & $1.53 \pm 0.5$                      & $8 \times 10^{-14}$                    & $5 \times 10^{-7}$                    & ...     \\ 
9838  &  54546.42945 & $1.54^{+0.41}_{-0.75}$    & $(8.7^{+3.7}_{-5.5}) \times 10^{-14}$    & $5 \times 10^{-7}$    & 0.50        \\ 
8990 &  54547.32918  & $2.14^{+0.36}_{-0.90}$    & $(6.0 \pm 5.3) \times 10^{-14}$          & $3 \times 10^{-7}$    & 0.57        \\ 
9839  &  54548.30002 & $1.30 \pm 0.78$           & $(4.6^{+4.3}_{-2.2}) \times 10^{-14}$    & $2 \times 10^{-7}$    & 0.34        \\ 
9837  &  54549.22579 & $3.15^{+0.40}_{-0.59}$    & $(1.6^{+0.5}_{-0.9}) \times 10^{-13}$    & $8 \times 10^{-7}$    & 0.27        \\ 
\multicolumn{2}{l}{\textit{MAXI J1659-152}} &    &     &        &    \\ 
12438                &  55665.96202    & $2.11 \pm 0.23$           & $(4.6 \pm 0.8) \times 10^{-13}$       & $2 \times 10^{-6}$          & 0.50        \\ 
12439               &  55674.74944     & $1.79 \pm 0.39$           & $(1.2 \pm 0.3) \times 10^{-13}$       & $4 \times 10^{-7}$    & 0.89        \\ 
12440\tablenotemark{e} &  55684.29844    & $2.26 \pm 0.5$                      & $6 \times 10^{-15}$                    & $2 \times 10^{-8}$                    & ...     \\ 
12441\tablenotemark{f} &  55693.21054     & $1.55^{+0.04}_{-0.02}$    & $(1.3 \pm 0.02) \times 10^{-11}$         & $4 \times 10^{-5}$         & 0.60        \\ 
12441\tablenotemark{g} &  55693.21054    & $1.74 \pm 0.33$    & $(1.2^{+0.1}_{-0.2}) \times 10^{-11}$    & $4 \times 10^{-5}$    & 0.89  \\ 
12442\tablenotemark{e} &  55788.83283   & $2.25 \pm 0.5$                      & $4 \times 10^{-15}$                    & $1 \times 10^{-8}$                    & ...     \\ 
12443                &  55846.53179  & $3.25 \pm 0.73$    & $(5.8^{+1.6}_{-1.2}) \times 10^{-15}$    & $2 \times 10^{-8}$    & 0.65        \\ 
\multicolumn{2}{l}{\textit{XTE J1752-223}} &    &     &        &    \\ 
11053                &  55389.63924   & $2.17 \pm 0.95$           & $(5.2^{+2.9}_{-1.9}) \times 10^{-14}$      & $1 \times 10^{-7}$    & 0.12        \\ 
12310                &  55397.07034   & $1.75 \pm 0.14$           & $(6.8 \pm 0.8) \times 10^{-13}$            & $2 \times 10^{-6}$    & 0.95        \\ 
11055                &  55403.24698   & $1.64^{+0.53}_{-0.86}$    & $(9.7 \pm 6.3) \times 10^{-15}$            & $2\times 10^{-8}$          & 0.36        \\ 
11056                &  55410.27430   & $2.20 \pm 0.47$           & $(1.4 \pm 0.5) \times 10^{-14}$            & $4 \times 10^{-8}$    & 0.13        \\ 
\hline
\end{tabular}
\\[0.1cm]

\tablenotetext{1}{Uncertainties are at the 68\% level ($\Delta C = 1.0$ for one parameter of interest, where $C$ is the Cash statistic).}
\tablenotetext{2}{Unabsorbed flux calculated by integrating over the best-fit model from 0.5--10 keV, excluding the effects of extinction.} 
\tablenotetext{3}{Normalized Eddington ratio $l_x = L_{0.5-10~keV}$/\ledd.}    
\tablenotetext{4}{A goodness value around 0.50 indicates a good fit to the data.  We fit an absorbed power law to 10$^4$ simulated spectra based on the best-fit parameters to each dataset.  The goodness is the fraction of fits with a lower Cash statistic than our best-fit to the real dataset.} 
\tablenotetext{5}{Effective photon indices and fluxes are estimated from X-ray band ratios using PIMMS} 
\tablenotetext{6}{Fit with \citet{davis01} pileup model has best-fit pileup-parameter $\alpha$=$0.12\pm0.11$ (estimated pileup fraction $f_{\rm pile} = 0.021$).} 
\tablenotetext{7}{Fit to readout streak with 201 source photons (see \S\ref{sec:pileup}).}  

\renewcommand\arraystretch{1}

\caption{Log of Other \textit{Chandra} X-ray observations}
\label{tab:litobslog}
\footnotesize
\begin{tabular}{l c c c c c c c}
 \hline \hline
                Obs.       		&  
                Obs.            &  
                MJD         		&  
                Time on             &   
                Net count rate                       &  
                Net source		&  
                Background 	&
                Ref.\tablenotemark{a}\\
	      ID	 			&  
	      date			& 
	      (d; UTC)			& 
	      source (ks)		& 
	      0.5--7 keV (counts s$^{-1}$)	 & 
	      counts             &  
	      counts               & 
	      \\
	      
\hline
\multicolumn{2}{l}{\textit{\avi}} &    &     &  &      &    \\ 
95     & 2000 Feb 29  &  51603.14750 &   42.1 & $(3.20 \pm 0.30) \times 10^{-3}$       & 134.7      & 2.3    & (1,2) \\ 
5479   & 2005 Aug 20  &  53602.35887 &   39.6 & $(8.04 \pm 0.48) \times 10^{-3}$    & 318.4      & 1.6   & (3)   \\ 
\multicolumn{2}{l}{\textit{GRO J1655-40}} &    &     &  &      &    \\ 
99     & 2000 Jul 01  &  51726.79797 &   42.6 & $(1.51 \pm 0.21) \times 10^{-3}$          & 64.1      & 0.9    & (1)  \\ 
10907  & 2009 Jun 08  &  54990.10230 &   18.2 & $(9.09 \pm 0.76) \times 10^{-3}$    & 165.5      & 0.5  & (4)    \\ 
\multicolumn{2}{l}{\textit{GX 339-4}} &    &     &  &      &    \\ 
4445   & 2003 Sep 29  &  52911.48765 &   28.3 & $(3.80 \pm 0.12) \times 10^{-2}$     & 1076.5     & 1.5   & (5)   \\ 
12410  & 2011 May 15  &  55696.67966 &   27.2 & $(2.63 \pm 0.10) \times 10^{-2}$    & 715.8      & 2.2   & (6)   \\ 
\multicolumn{2}{l}{\textit{V404 Cyg}} &    &     &  &      &    \\ 
97     & 2000 Apr 26  &  51660.68457 &   10.3 & $(1.64 \pm 0.04) \times 10^{-1}$      & 1689.8     & 1.2    & (1,7) \\ 
3808   & 2003 Jul 28  &  52848.86452 &   55.6 & $(3.48 \pm 0.08) \times 10^{-2}$    & 1933.9     & 4.1   & (7)  \\ 
\multicolumn{2}{l}{\textit{XTE J1118+480}} &    &     &  &      &    \\ 
3422   & 2002 Jan 12  &  52286.14907 &   45.8 & $(1.54 \pm 0.21) \times 10^{-3}$    & 70.7       & 1.3     & (2)  \\ 
\multicolumn{2}{l}{\textit{XTE J1550-564}} &    &     &  &      &    \\ 
1845   & 2000 Aug 21  &  51777.36325 &    5.1 & $(1.29 \pm 0.18) \times 10^{-2}$    & 65.8       & 0.2     & (8,1,9) \\ 
1846   & 2000 Sep 11  &  51798.20398 &    4.6 & $(2.41 \pm 0.25) \times 10^{-2}$    & 109.9      & 0.1     & (8,1,9) \\ 
3448   & 2002 Mar 11  &  52344.62513 &   26.1 & $(4.60 \pm 0.14) \times 10^{-2}$    & 1200.2     & 4.8    & (9)  \\ 
3672   & 2002 Jun 19  &  52444.37961 &   18.0 & $(3.14 \pm 0.48) \times 10^{-3}$    & 56.6       & 0.4    & (9)  \\ 
3807   & 2002 Sep 24  &  52541.83357 &   24.4 & $(8.74 \pm 0.64) \times 10^{-3}$    & 213.6      & 0.4  & (9)    \\ 
4368   & 2003 Jan 28  &  52667.18878 &   23.7 & $(1.03 \pm 0.07) \times 10^{-2}$    & 244.5      & 0.5   & (9)   \\ 
5190   & 2003 Oct 23  &  52935.30373 &   47.8 & $(3.03 \pm 0.27) \times 10^{-3}$    & 144.9      & 1.1   & (9)   \\ 
\multicolumn{2}{l}{\textit{XTE J1650-500}} &    &     &  &      &    \\
3400   & 2002 Jan 23  &  52297.99426 &   10.0 & $(2.76 \pm 0.05) \times 10^{-1}$    & 2764.5     & 23.5  & (10)   \\
3401   & 2002 Feb 04  &  52309.60891 &    9.5 & $(2.73 \pm 0.05) \times 10^{-1}$    & 2595.0     & 19.0    & (10)  \\
2731   & 2002 Mar 02  &  52335.09478 &   18.3 & $(5.02 \pm 0.05) \times 10^{-1}$    & 9177.9     & 10.1    & (10)  \\                                                                                                             
\hline
\end{tabular}
\\[0.1cm]
\tablenotetext{1}{References where data were previously published: (1) \citet{kong02}, (2) \citet{mcclintock03}, (3) \citet{gallo06}, (4) \citet{calvelo10}, (5) \citet{gallo03}, (6) \citet{corbel13}, (7) \citet{corbel08}, (8) \citet{tomsick01}, (9) \citet{corbel06}, (10) \citet{tomsick04}.} 
\end{table*}

\renewcommand\arraystretch{1.1}
\begin{table*}[htbp]
\center
\caption{Best-fit Spectral Parameters for Other  \textit{Chandra} X-ray observations}
\label{tab:litfits}
\centering
\begin{tabular}{l c c c  c c}
 \hline \hline
                Obs.      		&  
                MJD            &  
                $\Gamma$\tablenotemark{a}             &   
                Unabs.\ 0.5--10 keV flux\tablenotemark{a,b}&  
                $l_x$\tablenotemark{c}& 
                Goodness\tablenotemark{d} \\ 
	       ID	 			&  
	       (d; UTC)		& 
	      				& 
	      (erg cm$^{-2}$ s$^{-1}$)	 & 
	      	& 
	                       \\ 
\hline
\multicolumn{2}{l}{\textit{A 0620-00}} &       &      &    &    \\ 
95                   &  51603.14750  & $2.00 \pm 0.25$           & $(3.0 \pm 0.5) \times 10^{-14}$          & $4\times10^{-9}$    & 0.81        \\ 
5479                 &  53602.35887  & $2.24 \pm 0.16$           & $(8.9 \pm 0.9) \times 10^{-14}$          & $1\times10^{-8}$          & 0.32        \\ 
\multicolumn{2}{l}{\textit{GRO J1655-40}} &      &      &    &    \\ 
99                   &  51726.79797  & $1.78^{+0.45}_{-0.26}$    & $(3.9^{+0.8}_{-1.6}) \times 10^{-14}$    & $6 \times 10^{-8}$    & 0.72        \\ 
10907                &  54990.10230  & $1.93^{+0.13}_{-0.24}$    & $(2.1 \pm 0.5) \times 10^{-13}$    & $3 \times 10^{-7}$          & 0.58        \\ 
\multicolumn{2}{l}{\textit{GX 339-4}} &    &        &    &    \\ 
4445                 &  52911.48765   & $2.02 \pm 0.09$           & $(6.3^{+0.4}_{-0.3}) \times 10^{-13}$     & $7 \times 10^{-6}$    & 0.42        \\ 
12410                &  55696.67966   & $1.98 \pm 0.11$           & $(4.7 \pm 0.4) \times 10^{-13}$           & $5 \times 10^{-6}$          & 0.80        \\ 
\multicolumn{2}{l}{\textit{V404 Cyg}} &    &         &    &    \\ 
97                   &  51660.68457   & $2.00^{+0.04}_{-0.07}$    & $(3.2^{+0.1}_{-0.2}) \times 10^{-12}$    & $2\times10^{-6}$   & 0.15        \\ 
3808                 &  52848.86452   & $2.13^{+0.04}_{-0.07}$    & $(7.0 \pm 0.4) \times 10^{-13}$          & $5\times10^{-7}$    & 0.54        \\ 
\multicolumn{2}{l}{\textit{XTE J1118+480}} &      &      &    &    \\ 
3422                 &  52286.14907   & $2.09^{+0.34}_{-0.22}$    & $(9.7^{+0.8}_{-1.7}) \times 10^{-15}$    & $4\times10^{-9}$    & 0.56        \\ 
\multicolumn{2}{l}{\textit{XTE J1550-564}} &     &      &    &    \\ 
1845                 &  51777.36325   & $2.38^{+0.25}_{-0.52}$    & $(2.8^{+0.6}_{-0.9}) \times 10^{-13}$      & $8\times10^{-7}$    & 0.52        \\ 
1846                 &  51798.20398   & $2.21 \pm 0.39$           & $(5.7^{+0.9}_{-1.4}) \times 10^{-13}$      & $2\times10^{-6}$    & 0.53        \\ 
3448                 &  52344.62513   & $2.27^{+0.05}_{-0.08}$    & $(9.3^{+0.8}_{-0.3}) \times 10^{-13}$      & $3\times10^{-6}$    & 0.88        \\ 
3672                 &  52444.37961   & $2.57^{+0.25}_{-0.57}$    & $(7.8^{+1.8}_{-2.8}) \times 10^{-14}$      & $2\times10^{-7}$    & 0.55        \\ 
3807                 &  52541.83357   & $2.08^{+0.08}_{-0.15}$    & $(1.9 \pm 0.4) \times 10^{-13}$            & $6\times10^{-7}$          & 0.53        \\ 
4368                 &  52667.18878   & $2.14^{+0.15}_{-0.10}$    & $(2.4^{+0.2}_{-0.3}) \times 10^{-13}$      & $7\times10^{-7}$   & 0.64        \\ 
5190                 &  52935.30373   & $2.23 \pm 0.28$           & $(6.8 \pm 1.4) \times 10^{-14}$            & $2\times10^{-7}$    & 0.71        \\ 
\multicolumn{2}{l}{\textit{XTE J1650-500}}    &     &      &    &    \\
3400\tablenotemark{e} &  52297.99426 & $1.67 \pm 0.10$    & $(4.4^{+2.2}_{-0.9}) \times 10^{-11}$    & $4\times10^{-5}$    & 0.86        \\
3401\tablenotemark{f} &  52309.60891  & $1.64 \pm 0.10$           & $(4.3^{+4.9}_{-0.9}) \times 10^{-11}$    & $4\times10^{-5}$    & 0.82      \\
2731\tablenotemark{g} &  52335.09478  & $2.09 \pm 0.07$           & $(9.9 \pm 0.3) \times 10^{-12}$         & $9\times10^{-6}$          & 0.40        \\ 
\hline
\\[0.1cm]
\end{tabular}
\tablenotetext{1}{Uncertainties are at the 68\% level ($\Delta C = 1.0$ for one parameter of interest, where $C$ is the Cash statistic).}
\tablenotetext{2}{Unabsorbed flux calculated by integrating over the best-fit model from 0.5--10 keV, excluding the effects of extinction.}
\tablenotetext{3}{Normalized Eddington ratio $l_x = L_{0.5-10~keV}$/\ledd}	
\tablenotetext{4}{A goodness value around 0.50 indicates a good fit to the data.  We fit an absorbed power law to 10$^4$ simulated spectra based on the best-fit parameters to each dataset.  The goodness is the fraction of fits with a lower Cash statistic than our best-fit to the real dataset.}
\tablenotetext{5}{Fit with \citet{davis01} pileup model has best-fit pileup-parameter $\alpha$=$0.77 \pm 0.23$ (estimated pileup fraction $f_{\rm pile} = $0.34).} 
\tablenotetext{6}{Fit with \citet{davis01} pileup model has best-fit pileup-parameter $\alpha$=$0.73 \pm 0.27$ (estimated pileup fraction $f_{\rm pile} = $0.32). } 
\tablenotetext{7}{Fit with \citet{davis01} pileup model has best-fit pileup-parameter $\alpha$=$0.95_{-0.08}^{+0.05}$ (estimated pileup fraction $f_{\rm pile} = $0.098).} 
\end{table*}

\renewcommand\arraystretch{1}

\end{document}